\documentclass[reprint, amsmath, amssymb, aps]{revtex4-2}

\usepackage{float}
\usepackage{xcolor}
\usepackage{graphicx}
\usepackage{dcolumn}
\usepackage{bm}
\usepackage{epstopdf}

\PassOptionsToPackage{breaklinks}{hyperref}
\usepackage{xurl}
\usepackage{hyperref}

\hypersetup{%
  colorlinks=true,
  linkcolor=blue,
  citecolor=blue
}

\begin{document}

\title{A quantum algorithm for the linear Vlasov equation with collisions}

\author{Abtin Ameri}
\affiliation{Plasma Science and Fusion Center, Massachusetts Institute of Technology, Cambridge, MA 02139}
 \email{aameri@mit.edu}
\author{Paola Cappellaro}
\affiliation{ Research Laboratory of Electronics and Department of Nuclear Science and Engineering, Massachusetts Institute of Technology, Cambridge,
MA 02139
}%

\author{Hari Krovi}
\affiliation{
 Riverlane Research, Cambridge, MA 02142
}%
\author{Nuno F. Loureiro}
\affiliation{%
 Plasma Science and Fusion Center, Massachusetts Institute of Technology, Cambridge, MA 02139
}%

\author{Erika Ye}
\affiliation{
 Plasma Science and Fusion Center, Massachusetts Institute of Technology, Cambridge, MA 02139
}%

\date{\today}

\begin{abstract}

The Vlasov equation is a nonlinear partial differential equation that provides a first-principles description of the dynamics of plasmas. 
Its linear limit is routinely used in plasma physics to investigate plasma oscillations and stability. In this work, we present a quantum algorithm that simulates the linearized Vlasov equation with and without collisions, in the one-dimensional, electrostatic limit. Rather than solving this equation in its native spatial and velocity phase-space, we adopt an efficient representation in the dual space yielded by a Fourier-Hermite expansion. The Fourier-Hermite representation is exponentially more compact, thus yielding a classical algorithm that can match the performance of a previously proposed quantum algorithm for this problem. This representation results in a system of linear ordinary differential equations which can be solved with well-developed quantum algorithms: Hamiltonian simulation in the collisionless case, and quantum ODE solvers in the collisional case. In particular, we demonstrate that a quadratic speedup in system size is attainable.
\end{abstract}

\maketitle

\section{Introduction
\label{sec:intro}
}

Plasma dynamics is difficult to simulate with present-day computers due to the broad range of time- and length-scales that are typically exhibited by nonlinear plasma phenomena.
Indeed, direct numerical simulations covering scale ranges approaching those found in most real systems are beyond the capabilities of both current supercomputers and those predicted to exist in the near future. 
It is therefore natural to seek alternative computational platforms that can provide significant speedups. Quantum computers are a possible candidate since there exist quantum algorithms capable of outperforming their classical counterparts for a range of problems such as search via quantum walks~\cite{grover1996fast}, simulation of quantum mechanical systems (also called Hamiltonian simulation)~\cite{lloyd1996universal,berry2007efficient,childs2012hamiltonian,berry2014exponential,berry2015simulating,low2017optimal,low2019hamiltonian}, cryptography~\cite{shor1999polynomial}, and simulation of high-energy physics problems such as $1+1$ dimensional $\phi^4$ theory~\cite{JLP12}, fermionic field theory~\cite{fermionic_qft}, and conformal field theory~\cite{cft}. Furthermore, industrial and research efforts to commercialize quantum computers have significantly advanced the technology, bringing us closer to implementing quantum algorithms with real-world applications. 

Although quantum algorithms exist to simulate Hamiltonians and solve ordinary differential equations (ODEs), it is \textit{a priori} not clear if they can be used to solve specific plasma-physics problems, which are usually formulated in terms of coupled sets of nonlinear partial differential equations. Indeed, the field of designing quantum algorithms for nonlinear differential equations and, specifically, plasma physics, is still at an embryonic stage. So far, there has been a thorough exploration of quantum linear system algorithms (QLSAs)~\cite{harrow2009quantum,childs2017quantum}, linear ODEs~\cite{berry2014high,BCOW,childs2020,krovi2022improved}, certain regimes of nonlinear ODEs~\cite{leyton2008quantum,lloyd2020quantum,liu2021efficient,xue2021quantum,kyriienko2021solving}, and certain linear partial differential equations~\cite{cao2013quantum,costa2019quantum}. These algorithms use Hamiltonian simulation as a main subroutine. The application of such algorithms to real-world problems, such as plasma dynamics, is nontrivial and, in some cases, their promised speedup is lost. Nonetheless, there has been some progress in this field over the past few years. Dodin and Startsev~\cite{dodin2021applications} performed a survey of various approaches that could be used to solve plasma dynamics on quantum computers. Amongst those, they discussed the possibility of using the  Madelung transform to map the governing equations of a cold electron fluid to the Schr\"odinger equation. The Madelung transform was used by Zylberman \textit{et al}.~\cite{zylberman2022hybrid} to develop a hybrid algorithm simulating fluids. Linear embedding approaches have also been investigated, yielding a potential speedup if nonlinearities are weak~\cite{liu2021efficient,engel2021linear} or if the resulting system is sparse~\cite{joseph2020koopman,joseph2023quantum}. 
Other works have focused on developing quantum algorithms for wave problems in plasmas, namely the three-wave interaction~\cite{shi2021simulating} and extraordinary waves (an electromagnetic wave in a plasma where the wave's electric field is perpendicular to the background magnetic field) in cold plasmas~\cite{novikau2021quantum}. As one might expect, both these wave problems can be mapped to a Schr\"odinger-type equation, which can be solved using Hamiltonian simulation algorithms. A more general quantum lattice representation for one-dimensional wave propagation in plasmas has also been developed~\cite{vahala2022quantum} (but such an algorithm requires four qubits per node).

Ideally, one would like to have a quantum algorithm to solve a first-principles description of plasma physics as offered by the Vlasov equation. A first step in this direction is to understand if the linear limit of the Vlasov equation can be simulated efficiently on a quantum computer. This problem was explored by Engel, Smith, and Parker (henceforth referred to as ESP), where a quantum algorithm was developed that simulates the Landau damping of Langmuir waves~\cite{engel2019quantum}, a quintessential plasma physics phenomenon whereby electrostatic waves in the plasma resonantly transfer energy to electrons and are thus damped~\cite{landau196561}. By linearizing the collisionless Vlasov equation coupled to Amp\`ere's law, ESP mapped the problem to Hamiltonian simulation which can then be solved using recent algorithms~\cite{engel2019quantum}. Using a spatial grid of $N_v$ points, ESP gave a quantum algorithm to simulate the linearized Vlasov evolution with gate complexity of $\mathcal{O}(\mathrm{polylog} N_v\log (1/\epsilon))$, with $\epsilon$ being the error in the norm of the solution state. To extract the Landau damping rate, the electric field at different times must be obtained by sampling the final state (so as to obtain a time history from which the Landau damping rate can be computed in post-processing). The number of times one needs to repeat the algorithm is independent of $N_v$ and depends only on the absolute precision $\delta$ with which one needs to calculate the electric field, leading to overall run time of $\mathcal{O}((1/\delta)\mathrm{polylog} N_v)$, where $\delta \sim \epsilon$.

In this paper, we revisit this problem with a different mathematical approach. We find that, using the linearized Vlasov equation coupled to Poisson's equation and expanding the distribution function in velocity space using $M+1$ Hermite polynomials (with $N=M+1$ corresponding to the system size), we can obtain a system matrix which is exponentially smaller than that of ESP for the same precision $\epsilon$. Thus, a classical algorithm solving this system can be as efficient as the quantum algorithm of ESP. Furthermore, the Fourier-Hermite representation allows for the inclusion of collisions without changing the system matrix structure. We also find that a quantum algorithm using this framework yields a quadratic speed up compared to classical ODE solvers, which are the fastest classical algorithms for solving this problem~\cite{burden2015numerical}.

This paper is structured as follows. Section~\ref{sec:background} provides a brief background on the Vlasov equation, Landau damping and the Hermite representation, and rigorously defines the problem we aim to solve in addition to its key parameters. Section~\ref{sec:quantum} discusses the performance of a quantum algorithm based on the Hermite formalism for both collisionless and collisional cases, and demonstrates that a quadratic speedup is possible. Section~\ref{sec:erroranalysis} reviews the ESP formulation and performs a comparative error analysis of that system and ours. Finally, Section~\ref{sec:discon} summarizes our results and discusses their implications.

\section{Background
\label{sec:background}
}
\subsection{The Vlasov Equation}
The collisional Vlasov-Poisson system is given by
\begin{gather}
\label{eq:vlasovnonlinear}
    \frac{\partial f_s}{\partial t} + \mathbf{v} \cdot \nabla f_s + \frac{q_s}{m_s} \mathbf{E} \cdot \nabla_v f_s = C[f_s], \\
    \label{eq:poisson0}
    \epsilon_0 \nabla \cdot \mathbf{E} = \rho,
\end{gather}
where $\epsilon_0$ is the permittivity of free space, $s$ denotes a species (ions or electrons) of charge $q_s$ and mass $m_s$, and $\rho$ denotes the charge density. 
The collision operator is represented by $C[\cdot]$ and $\mathbf{E}$ is the electric field. We are going to restrict ourselves to the electrostatic limit, so $\mathbf{E}=-\nabla\varphi$, where $\varphi$ is the electrostatic potential.

The Vlasov-Poisson system describes the time evolution of the species' distribution function $f_s(\mathbf{x},\mathbf{v},t)$ in three-dimensional physical space and three-dimensional velocity space. The distribution function is defined such that the number of particles per unit volume in the vicinity of position $\mathbf{x}$ and velocity $\mathbf{v}$ is $f_s(\mathbf{x},\mathbf{v},t) d^3\mathbf{x}d^3 \mathbf{v}$. It is normalized such that
\begin{equation}
    \int f_s(\mathbf{x},\mathbf{v},t) d^3 \mathbf{v} = n_s(\mathbf{x},t),
\end{equation}
where $n_s$ is the species' number density. Therefore, the charge density is obtained from the distribution functions as
\begin{equation}
    \rho = \sum_{s = i,e} q_s \int f_s d^3 \mathbf{v}.
\end{equation}
For simplicity, in this work we will limit ourselves to one spatial and one velocity dimensions--$z$ and $v$, respectively--as in the textbook formulation of the Landau damping problem~\cite{chen2012introduction}. In this case, Eqs.~(\ref{eq:vlasovnonlinear},\ref{eq:poisson0}) become
\begin{gather}
    \frac{\partial f_s}{\partial t} + v \frac{\partial f_s}{\partial z} + \frac{q_s}{m_s} E \frac{\partial f_s}{\partial v} = C[f_s], \label{eq:nonlinearVlasov}\\
    \epsilon_0 \frac{\partial E}{\partial z} = \sum_{s = i,e} q_s \int f_s d v.\label{eq:nonlinearpoisson}
\end{gather}
We can linearize Eqs.~(\ref{eq:nonlinearVlasov}-\ref{eq:nonlinearpoisson}) about a Maxwellian background, and consider the case where the ions are stationary, meaning that their distribution function is not perturbed, so we need only consider the evolution of the electron distribution function. 
Then, $f \approx F_{0} + g$, where we have dropped the species subscript for simplicity, $g \ll F_{0}$ is the perturbation to the distribution function, and $F_{0}$ is the (one-dimensional) Maxwellian distribution function:
\begin{equation}
    F_{0}(v) =   \frac{1}{\sqrt{\pi}} \frac{n_0}{v_{th}}e^{-v^2/v_{th}^2},
\end{equation}
where $n_0$ is the background density and
\begin{equation}
    v_{th} = \sqrt{\frac{2k_B T}{m}},
\end{equation}
is the electron thermal velocity, with $T$ and $k_B$ denoting the temperature and Boltzmann constant, respectively. We carry out the linearization by noting that $C[F_{0}] = 0$, ignoring quadratic perturbative terms, and using the normalizations 
\begin{equation}
    \begin{aligned}
    z &\leftarrow \frac{z}{\sqrt{2} \lambda_{D}}, \\
    t &\leftarrow \omega_{p} t, \\
    v &\leftarrow \frac{v}{v_{th}}, \\
    f &\leftarrow \frac{v_{th}}{n} f, \\
    \varphi &\leftarrow \frac{e}{k_B T} \varphi,
\end{aligned}
\end{equation}
where $e$ is the fundamental charge and
\begin{equation}
    \omega_{p} = \sqrt{\frac{n_0 e^2}{\epsilon_0 m}}, \ \ \ \ \lambda_{D} = \sqrt{\frac{\epsilon_0 k_B T}{n_0 e^2}},
\end{equation}
are, respectively, the plasma frequency and the Debye length. Thus, the linearized Vlasov-Poisson system becomes
\begin{gather}
\frac{\partial g_k}{\partial t} + ikvg_k + ikv F_0\varphi_k =  C[g_k], \label{eq:linvlasov}  \\
\varphi_k = \alpha\int_{-\infty}^{+\infty} g_k dv, \label{eq:poisson}
\end{gather}
where we have also Fourier-transformed both equations, with $k$ representing the Fourier wavenumber, and $\alpha =  2/k^2$ is the physics parameter. 
By choosing a different value of $\alpha$ satisfying $\alpha \geq -1$, different target problems can be described, where Eqs.~(\ref{eq:linvlasov}-\ref{eq:poisson}) consider the evolution of only one species, and $\alpha$ prescribes the behavior of the other species (Boltzmann, isothermal, or no response)~\cite{kanekar2015fluctuation,adkins2018solvable}.

While the Vlasov equation is typically coupled to Poisson's equation in the electrostatic case, it can also be coupled to Amp\`ere's law:
\begin{equation}
    \frac{\partial E_k}{\partial t} = -J_k,
\end{equation}
where $J_k$ is the $k$-th Fourier mode of the current. For the case of Langmuir waves, the linearized Amp\`ere's law, coupled to the Vlasov equation in Fourier space, yields the following set of equations:
\begin{gather}
    \frac{\partial g_k}{\partial t} + ikvg_k - v F_0E_k =  C[g_k], \label{eq:linvlasovESP} \\
    \frac{\partial E_k}{\partial t} = \int v g_k \ dv. \label{eq:ampereESP}
\end{gather}
While Eqs.~(\ref{eq:linvlasov}-\ref{eq:poisson}) and (\ref{eq:linvlasovESP}-\ref{eq:ampereESP}) are two different formulations, they describe the same physics.

\subsection{Landau Damping} \label{sec:landaudamping}
Landau damping is a textbook plasma phenomenon whereby wave energy is resonantly transferred to plasma particles.
Despite what its name might suggest, Landau damping is a reversible process: in the absence of collisions, the total energy in a plasma is conserved throughout this process. During Landau damping, the plasma distribution function attains continuously finer structures in velocity space. In a collisionless plasma, these structures continue to develop until the end of the linear regime. In a collisional plasma, such structures are eventually smoothed out by collisions. This will become important when we perform the error analysis in Section~\ref{sec:erroranalysis}.

\begin{figure}[h!]
    \centering
    \includegraphics[width=8.6 cm]{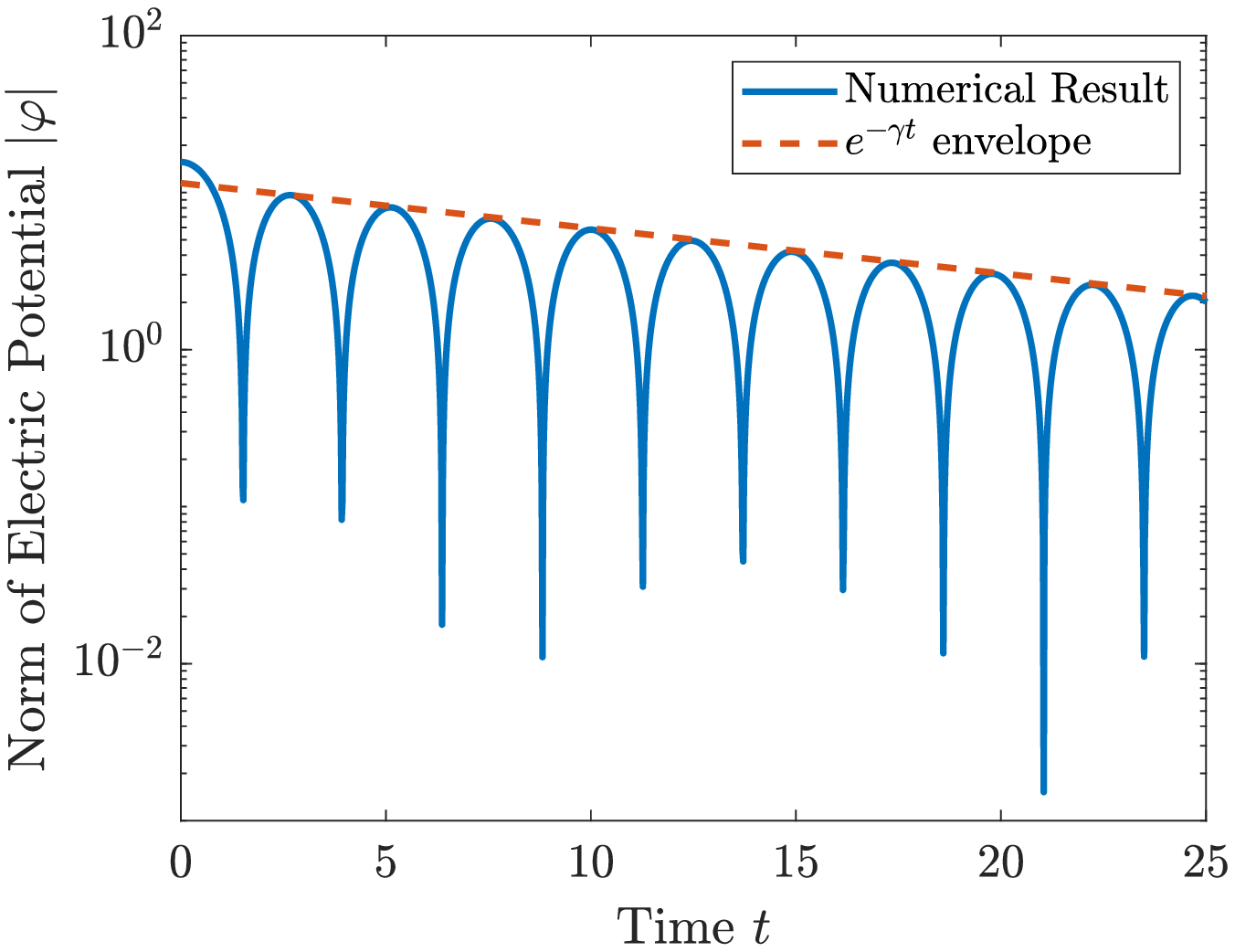}
    \caption{The norm of the electric potential $|\varphi|$ as a function of time $t$. The blue solid line is obtained by numerically solving Eqs.~(\ref{eq:linvlasov}-\ref{eq:poisson}) with $k = 0.56569$, and $\alpha = 2/k^2$. The red dashed line is an $e^{-\gamma t}$ envelope, where $\gamma = 0.066237$.}
    \label{fig:landaudamping}
\end{figure}

To calculate the Landau-damping rate of a plasma wave, one typically tracks the time evolution of the amplitude of the electric potential (or, equivalently, the electric field) in the plasma. The amplitude can be fitted by an $e^{-\gamma t}$ envelope, where $\gamma > 0$ is the Landau damping rate. 
The density perturbation in a plasma can also be used to calculate the Landau damping rate, since the electric potential and density perturbation are proportional to each other, as per Eq. (\ref{eq:poisson}). This procedure is illustrated in Fig.~(\ref{fig:landaudamping}), where the time evolution of the norm of the electric potential $|\varphi|$ is tracked. The peaks are fitted with an $e^{-\gamma t}$ envelope, and the correct Landau damping rate is recovered from the fit.

\subsection{The Hermite Representation}
Working with the linear Vlasov-Poisson system, it is often convenient, and numerically advantageous, to express the distribution function as a series of Hermite polynomials~\cite{armstrong1967numerical,grant1967fourier,armstrong1970solution,hammett1993developments,parker1995renormalized,sugama2001collisionless,zocco2011reduced,kanekar2015fluctuation}:
\begin{equation} \label{eq:hermiteexpansion}
g_k(v,t) = \sum_{m=0}^\infty \frac{H_m(v) F_0(v)}{\sqrt{2^m m!}
}g_{m,k}(t),
\end{equation}
where $H_m(v)$ denotes the Hermite polynomial of order $m$ and $g_{m,k}(t)$ is its coefficient given by
\begin{equation}
    g_{m,k}(t) = \int_{-\infty}^{+\infty} dv \frac{H_m(v)}{\sqrt{2^m m!}}g_k(v,t).
\end{equation}
Adopting a Lenard-Bernstein-type collision operator for simplicity~\cite{lenard1958plasma},

\begin{equation}
    C[g_k] = \nu \frac{\partial}{\partial v} \left ( \frac{1}{2} \frac{\partial}{\partial v} + v \right )g_k,
\end{equation}
where $\nu \geq 0$ is the collision frequency, and introducing new variables defined as
\begin{align}
    \tilde{g}_{0,k} &= g_{0,k}, \\
    \tilde{g}_{m,k} &= \frac{1}{\sqrt{1+\alpha}}g_{m,k}, \ m\geq 1 ,
\end{align}
the Hermite-transformed equations become a coupled set of linear advection equations for the Hermite-Fourier coefficients, as follows: 
\begin{align} 
\label{eq:hermitesystemexpanded_g0}
    &\frac{d g_{0,k}}{d t} + ik \sqrt{\frac{1+\alpha}{2}}g_{1,k}=0,\\
    \label{eq:hermitesystemexpanded_g1}
   & \frac{d g_{1,k}}{d t} + ik\left(g_{2,k}+\sqrt{\frac{1+\alpha}{2}}g_{0,k}\right)=0, \\
    \label{eq:hermitesystemexpanded_gm}
    &\frac{d g_{m,k}}{d t} + ik \left ( \sqrt{\frac{m+1}{2}}g_{m+1,k} + \sqrt{\frac{m}{2}}g_{m-1,k} \right)\nonumber\\
    &= -\nu m g_{m,k},\ m\ge 2,
\end{align}
where tildes have been dropped for notational simplicity.
Provided that an infinite number of Hermite moments are kept, these equations are formally equivalent to Eqs.~(\ref{eq:linvlasov}-\ref{eq:poisson}). 
For practical purposes, however, only a finite number of moments, $M+1$, must be retained. Since the equation for the Hermite moment of order $m$ couples to the successive moment, $m+1$, a truncation is needed to integrate these equations on a computer. We adopt the customary closure $g_{M+1}=0$, yielding the following system
\begin{align} \label{eq:hermitetransformed}
    \frac{d \mathbf{g}_k}{dt} = A\mathbf{g}_k,
\end{align}
with solution
\begin{align}
    \mathbf{g}_k(t) = e^{At} \mathbf{g}_k(t=0),
\end{align}
where $\mathbf{g}_k = [g_{0,k} \cdots g_{M,k}]^T$, $\mathbf{g}_k(t=0)$ is the initial condition, and $A$ is a 3-sparse matrix containing the relevant coefficients,
\begin{equation}\label{eq:A}
    A=-iH+\nu\,\mathrm{diag}([0,0,2,\dots,M-1,M]),
\end{equation}
with
\begin{equation} \label{eq:H}
    H =k
        \begin{bmatrix}
            0 &  \sqrt{\frac{1+\alpha}{2}} &  &   &  &  \\
             \sqrt{\frac{1+\alpha}{2}} & 0 & 1 &   & &  \\
            & 1 & 0 &  \sqrt{\frac{3}{2}}  &   &  \\
             &  & \ddots & \ddots & \ddots &   \\
             &  &  &   \sqrt{\frac{M - 1}{2}} & 0 &  \sqrt{\frac{M}{2}} \\
             &  &  &   &   \sqrt{\frac{M}{2}} & 0
        \end{bmatrix}.
\end{equation}

The initial condition for the canonical Landau damping problem is $g_0(z,v,t=0)=\bar g_0 \cos(k_0 z)$, where $\bar g_0$ is a constant (an arbitrary amplitude). In Fourier-Hermite space, this initial condition is
\begin{equation} \label{eq:initialcondition}
   g_{m,k}(t=0) = \delta_{m,0}(\delta_{k,k_0} + \delta_{k,-k_0}),
\end{equation}
where we have chosen $\bar g_0 = \sqrt{2/\pi}$.

While $M$ is in principle a free parameter, typically one wishes to have $M \gtrsim \omega/\nu$ in order to capture the physics, where $\omega$ is a characteristic frequency of the system, to ensure that the higher Hermite moments are properly dissipated by collisions. If $\nu = 0$, $M$ will be a function of simulation time $T$. This will be relevant to the discussion in Section~\ref{sec:erroranalysis}.

\section{\label{sec:quantum}The Quantum Algorithm}
In this Section~we will show that a quantum algorithm solving Eq.~(\ref{eq:hermitetransformed}) yields a quadratic speedup with respect to system size compared to the most efficient classical algorithms.

\subsection{Collisionless Case and Hamiltonian Simulation}
\begin{figure}[h!]
    \centering
    \includegraphics[width=8.6 cm]{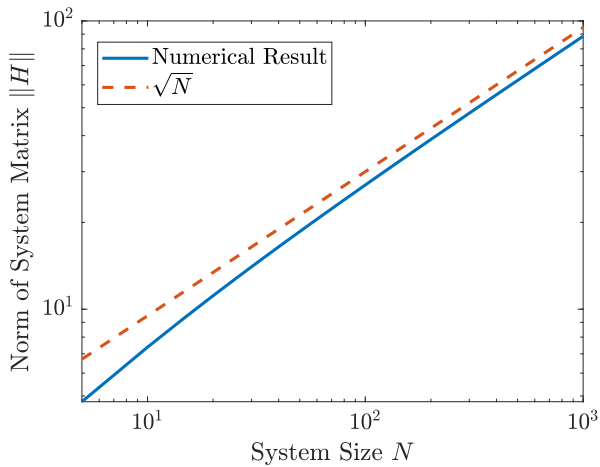}
    \caption{Numerically obtained norm of the system matrix $\|H\|$ (Eq.~(\ref{eq:H})) as a function of the system size $N=M+1$, where $M$ is the number of Hermite moments. The orange dashed line is a $\sqrt{N}$ scaling.}
    \label{fig:system_norm_collisionless}
\end{figure}

When $\nu = 0$ we have a collisionless system where $A=-iH$ with $H$ a Hermitian matrix. Such a system conserves total energy and, thus, is governed by unitary time evolution generated by $H$:

\begin{align}
    \frac{d \mathbf{g}_k}{dt} = -iH \mathbf{g}_k,
\end{align}
Since this is equivalent to a Schr\"odinger equation, it can be simulated on a quantum computer using Hamiltonian simulation techniques. An efficient quantum algorithm for simulating an $s$-sparse Hamiltonian has query complexity $\mathcal{O}\left (s\|H\|T + \frac{\log (1/\epsilon)}{\log \log (1/\epsilon)} \right)$~\cite{low2017optimal}, where $s$ is the sparsity, $\|H\|$ is the norm, $T$ is the final simulation time, and $\epsilon$ is the error in the norm of the solution state. In the Hermite system, the system matrix has sparsity $s=2$. Furthermore, the norm of the system scales with the system size $N = M+1$ as
\begin{equation}
    \|H\| \sim \mathcal{O} (\sqrt{N}).
\end{equation}
This is also shown in Fig. \ref{fig:system_norm_collisionless}, where we plot the norm of the system matrix $\|H\|$ as a function of system size $N$ and observe the $\mathcal{O}(\sqrt{N})$ scaling.

\subsection{Collisional Case and Quantum ODE Solvers}
For the more realistic scenario where $\nu \neq 0$, unitary time evolution is lost due to the presence of the collision terms. In this case, we need to solve an ODE rather than simulate a Hamiltonian. We use the results from Ref.~\cite{krovi2022improved}, which state the following. Suppose $A$ is an $s$-sparse matrix with oracle access to its entries. Then Eq.~(\ref{eq:hermitetransformed}) can be solved to produce a quantum state proportional to the solution in time $O(\mathcal{G}\,T\,\|A\|\, \mathcal{C}(A))$, where $\mathcal{G} = \mathrm{max}_{t \in [0,T]} \|\mathbf{g}_k(t)\|/\|\mathbf{g}_k(T)\|$ measures the solution decay and $\mathcal{C}(A) = \mathrm{sup}_{t \in [0,T]} \|\mathrm{exp}(At)\|$. For our system, the matrix $A$ has negative log norm and so $\mathcal{C}(A)\leq 1$. Furthermore, since $M \sim \omega/\nu$, the addition of collisions does not change the $\mathcal{O}(\sqrt{N})$ scaling of $\|A\|$ as long as $\nu \ll k$. To see this, we can approximate $\omega$ by choosing $M=2$ and solving for the dispersion relation, giving us
\begin{equation}
    \omega = \sqrt{\frac{1+\alpha}{2}}k.
\end{equation}
Typically, $\alpha = \mathcal{O}(1)$, so $\omega \sim k$. To maintain the $\mathcal{O}(\sqrt{N})$ scaling of $\|A\|$, we demand that the off-diagonal terms of Eq.~(\ref{eq:A}) dominate the diagonal terms, yielding the following constraint:

\begin{equation}
    \nu \sqrt{M} \ll k.
\end{equation}
Knowing that $M\sim \omega/\nu \sim k/\nu$, we obtain $\nu \ll k$. This is satisfied in most plasmas of interest because $\nu$, which is normalized by the plasma frequency $\omega_p$, is very small compared to $k$, which is normalized by Debye length $\lambda_D$. Thus, even in the presence of collisions, we can still obtain a quadratic speedup in system size. However, the time dependence of the algorithm suffers. This is because collisions cause the norm of the solution to decay exponentially, meaning that for large $T$, $\|\mathbf{g}_k(0)\|/\|\mathbf{g}_k(T)\|$ scales as $e^{\nu T}$. To obtain $\mathbf{g}_k(T)$ efficiently, we can restrict the simulation time, which we will discuss in Section~\ref{sec:output}.

\subsection{Oracle Representation, State Preparation, and Output} \label{sec:output}
The Hamiltonian simulation and quantum ODE solvers require access to the system matrix $A$ through an oracle $O_A$. The oracle has the following action:
\begin{equation}
    O_A |i,j,0\rangle = |i,j,A_{ij}\rangle,
\end{equation}
where $A_{ij}$ denotes the element in the $i$-th row and $j$-th column of $A$. The representation of $O_A$ is as follows:
\begin{equation}
    \begin{aligned}
    O_A &=  \left ( |1,2 \rangle \langle 1,2| + |2,1\rangle \langle 2,1| \right ) \otimes \left |-ik\sqrt{\frac{1+\alpha}{2}} \right\rangle \langle 0| \\
    &+ \sum_{l = 2}^{M} |l,l+1\rangle \langle l,l+1| \otimes \left|-ik \sqrt{\frac{l}{2}}\right\rangle \langle 0| \\
    &+ \sum_{l = 3}^{M+1} |l,l-1\rangle \langle l,l-1| \otimes \left|-ik \sqrt{\frac{l-1}{2}}\right\rangle  \langle 0|\\
    &+ \sum_{l = 3}^{M+1} |l,l\rangle \langle l,l| \otimes |(l-1)\nu\rangle  \langle 0|.
\end{aligned}
\end{equation}
This oracle can be block-encoded using the methods in~\cite{sunderhauf2023block}.

The initial condition for the Hermite system is given by Eq.~(\ref{eq:initialcondition}). This is trivial to construct, and does not add to the complexity of the algorithm.

 In general, one has access to all Hermite moments in the solution state. If one is interested only in computing the Landau damping rate, then it suffices to measure the zeroth Hermite moment $g_{0,k}$, from which the electric potential can be calculated
\begin{equation}
    \varphi_k = \alpha g_{0,k}.
\end{equation}
To obtain the Landau damping rate, $g_{0,k}$ needs to be sampled at various times. To obtain $g_{0,k}$ at a particular time, amplitude estimation can be used~\cite{brassard2002quantum}. This requires $\mathcal{O}(1/\delta)$ repetitions of the algorithm to calculate $g_{0,k}$ to absolute precision $\delta$. Once the amplitude is obtained, the algorithm must be repeated for the other sample times. Doing amplitude estimation efficiently at all times is contingent upon the electric potential decaying slowly. Furthermore, in the collisional case, the simulation time must be limited due to the solution norm decaying as $e^{-\nu t}$. In general, one is interested in kinetic effects, which are captured when $\nu \ll \gamma$. This means that the collisions are occurring on a timescale which is longer than the timescales of interest (Landau damping, in this case). Thus, by choosing $T\sim 1/\gamma$, we can ensure that we recover the Landau damping rate accurately without the decay of both the solution norm and the electric potential hindering the performance.

These repetitions required for extracting the Landau damping rate increase the complexity of the quantum algorithm with respect to error. Efficient Hamiltonian simulation algorithms to prepare the solution state scale as $\mathcal{O}(\log (1/\epsilon))$, but extracting a classical parameter (such as the Landau damping rate) by using amplitude estimation -- assuming that $\delta \sim \epsilon$ -- turns this scaling into $\mathcal{O}(1/\epsilon)$.

\section{\label{sec:erroranalysis}Error Analysis}

\subsection{\label{sec:problemstatement} Statement of the Problem}
Given a precision $\epsilon$, we would like to determine the system size $N$ required to obtain $g_{0,k}$ from~Eq.~(\ref{eq:hermitetransformed}), as well as the system size 
 $N_v$ required to obtain $E_k$ from Eqs.~(\ref{eq:linvlasovESP}-\ref{eq:ampereESP}). 
In the former case, the input parameters are the simulation time $T$~\footnote{The linear Vlasov equation exhibits a phenomenon known as recursion, where, for a long simulation time, the velocity space cascade towards finer structures reverses. This phenomenon is purely numerical and requires one to choose the velocity space resolution to be high enough (or the simulation time to be small enough) to prevent it. For the error analysis, we will assume that this is indeed the case, so as to avoid the problem of recursion.}, the Fourier coefficient $k_0$ for the initial condition, the collision frequency $\nu$, and physics parameter $\alpha$. 
For the ESP system, in addition to $T$ and $k_0$, the velocity grid truncation value $v_{max}$ is required (their algorithm only solves the collisionless case, $\nu=0$).

\subsection{\label{sec:level2} The Hermite System}

The error analysis for the Hermite system is non-trivial and depends on the function that is being Hermite-expanded. Typically, for the types of functions involved in this problem (decaying as $e^{-v^2}$), the convergence of the Hermite series is super-exponential~\cite{boyd1984asymptotic,boyd2001chebyshev}. That is,
\begin{equation}
    \epsilon = \Omega\left ( e^{-\beta M} \right ), 
\end{equation}
where $\beta = \beta (k,T)$. 

\begin{figure} [h]
    \centering
    \includegraphics[width=8.6 cm]{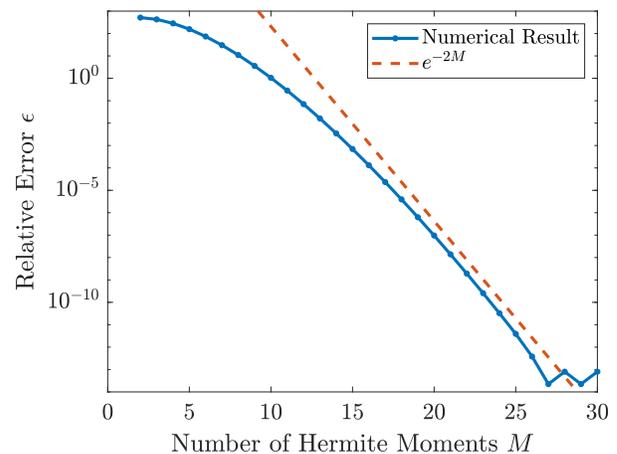}
    \caption{The relative error $\epsilon$ as a function of the number of Hermite moments $M$. The blue solid line is the error obtained from solving Eq.~(\ref{eq:hermitetransformed}) using an ODE solver with $\alpha = \nu = 0$, $k = 2$, and $T = 2.5$, and the dashed orange line is an $e^{-2M}$ scaling.}
    \label{fig:hermite_convergence}
\end{figure}

Since an exact analytical solution to Eqs.~(\ref{eq:linvlasov}-\ref{eq:poisson}) is not known, 
for the purpose of the error analysis we analyze the case $\alpha = 0$ (as done in other works, e.g.~\cite{armstrong1970solution,hammett1993developments}), which corresponds to the free-streaming equation
\begin{equation}
    \frac{\partial g_k}{\partial t} + ikvg_k = 0,
\end{equation}
with initial condition
\begin{equation}
    g_k(v,0) = e^{-v^2},
\end{equation}
whose exact solution is
\begin{equation} \label{eq:esp_exact}
    g_k (v,t) = e^{-ikvt-v^2}.
\end{equation}
The Hermite coefficients of this solution which solve Eq.~(\ref{eq:hermitetransformed}) are given by

\begin{equation} \label{eq:exact}
    g_{m,k}(t) = \frac{e^{-k^2t^2/4} (-ikt)^m}{\sqrt{2^m m!}}.
\end{equation}
The factorial in the denominator scales as $\mathcal{O}(e^{m \log m})$ and results in a super-exponential convergence as $m$ increases. Fig.~\ref{fig:hermite_convergence} plots the convergence of the Hermite series for the zeroth moment $g_{0,k}$. An ODE solver is used to solve Eq.~(\ref{eq:hermitetransformed}) and the solution is compared with Eq.~(\ref{eq:exact}) to obtain the relative error. The figure shows the predicted convergence.

This exponential convergence can also be demonstrated in the collisional case. Indeed, as shown in Ref.~\cite{zocco2011reduced}, seeking a steady-state solution to Eq.~(\ref{eq:hermitesystemexpanded_gm}) results in a Hermite spectrum given by
\begin{equation}
    |g_m| = \mathcal{O}\left ( \frac{1}{m^{1/4}} \exp \left[-\frac{1}{2}\left (\frac{m}{m_c} \right)^{3/2} \right]  \right ),
\end{equation}
where $m_c$ corresponds to the collisional cutoff (a function of the collision frequency and other physical inputs of the problem). While the convergence is polynomial with $1/m$ for small values of $m$, it becomes exponential asymptotically for larger values of $m$.

The fast convergence means that the system size $N=M+1$ scales logarithmically with the error:
\begin{equation}
    N = \mathcal{O} \left ( \log (1/\epsilon) \right ). \label{eq:scaling_hermite}
\end{equation}

\subsection{\label{sec:ESP}The ESP System}
Here we will briefly overview the approach of ESP~\cite{engel2019quantum} and perform an error analysis. The reader is referred to their paper for further details. 

Instead of using the Poisson equation, as we do in our formulation, ESP couple the linear (collisionless) Vlasov equation to Amp\`ere's law, as in Eqs.~(\ref{eq:linvlasovESP}-\ref{eq:ampereESP}) with $C[g_k]=0$. They then restrict velocity space to the domain $[-v_{max},v_{max}]$, where $v_{max}$ is an input parameter, and approximate the integral in Eq.~(\ref{eq:ampereESP}) using a Riemann sum on a velocity grid of $N_v$ points. This, along with variable transformations, allows them to write the system as a Schr\"odinger-type equation,

\begin{equation}
    \frac{d\boldsymbol{\psi}_k}{dt} = -iH\boldsymbol{\psi}_k,
\end{equation}
where $\boldsymbol{\psi}_k$ stores the (Fourier-transformed) distribution function in velocity space, as well as the amplitude of the electric field. 

\begin{figure} [h]
    \centering
    \includegraphics[width=8.6 cm]{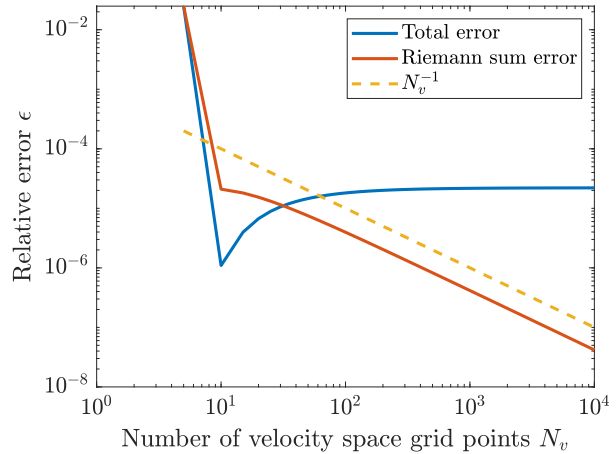}
    \caption{The relative error in integrating $\int_{-\infty}^{\infty} e^{-v^2} \ dv$ as a function of the number of velocity space grid points $N_v$ with $v_{max}=3$. The blue solid line is the total error, as shown in Eq.~(\ref{eq:esperror}), the orange line is the Riemann sum error, as shown by the second term on the right-hand-side of Eq.~(\ref{eq:esperror}), and the dashed yellow line is an $N_v^{-1}$ scaling.}
    \label{fig:esp_convergence}
\end{figure}

The error in the amplitude of the electric field originating from Eq.~(\ref{eq:ampereESP}) is given by
\begin{equation}
    \epsilon = \mathcal{O} \left ( e^{-v_{max}^2} + \frac{2L_1v_{max}^2}{N_v} \right ) \label{eq:esperror},
\end{equation}
where $L_1 = \max_{v\in [-v_{max},v_{max}]} |dg_k/dv|$. The first term corresponds to the domain truncation error~\cite{boyd2001chebyshev}, and the second corresponds to the Riemann sum error~\cite{rall1965numerical}. Because $L_1$ is dependent on the amount of structure present in the perturbed distribution function, we can conclude that $L_1 = L_1 (k,T)$ (as per the discussion in Section~\ref{sec:landaudamping}). As an example, Fig. \ref{fig:esp_convergence} is a convergence plot for approximating $\int_{-\infty}^{\infty} e^{-v^2} \ dv$ as a domain-truncated Riemann sum, showing convergence of the error with respect to $N_v$. The Gaussian function is specifically chosen for the error analysis because $g_k$ decays as a Gaussian as $|v| \rightarrow \infty$. We can see the total error and the Riemann sum error, and we observe that the latter converges as $N_v^{-1}$, as predicted. The saturation of the total error as $N_v$ increases is due to the domain truncation error (the difference between the blue and orange lines).

For simplicity, let us take the domain truncation error and the Riemann sum error to be of the same order:
\begin{equation}
     e^{-v_{max}^2} \sim \frac{2L_1v_{max}^2}{N_v} \sim \epsilon.
\end{equation}
Then, to achieve a precision $\epsilon$, we have the following expression for $N_v$:
\begin{equation}
    N_v = \mathcal{O} \left ( \frac{\log (1/\epsilon)}{\epsilon} \right). \label{eq:scaling_esp}
\end{equation}
Thus, the ESP system size $N_v$ scales polynomially with $1/\epsilon$.

\subsection{Comparison}

Comparing Eqs.~(\ref{eq:scaling_hermite}) and (\ref{eq:scaling_esp}), we see that the Hermite system can achieve the same precision $\epsilon$ with an exponentially smaller system size $N$ compared to the ESP system. This means that a classical ODE solver -- whose complexity scales as $\mathcal{O}(N)$ --  implementing the Hermite system has the same performance as the quantum algorithm of ESP.

While we have excluded the dependence of the system size on the simulation time $T$ in the error analysis, we note that this dependence is different for the Hermite and ESP representations. Inspecting Eq.~(\ref{eq:esp_exact}), we find that structures develop from the $\exp (-ikTv)$ term, meaning that the ESP representation would need $N_v \sim kT$ grid points to resolve such structures. On the other hand, from Eq.~(\ref{eq:exact}) we find that the peak of $|g_{M,k}(T)|$ occurs when $M = (kT)^2/2$; this sets the minimum resolution required to capture the physics. Thus, in the long simulation time limit, the ESP system size has better dependence on $T$. However, this does not affect the results of this paper. This is because the ESP system size's dependence on $T$ and $\epsilon$ goes as $N_v \sim T \log(1/\epsilon)/\epsilon$, whereas for the Hermite system size the dependence is $N \sim T^2 \log (1/\epsilon)$. At first glance, it may seem that, by fixing $T$, the Hermite system size would be quadratically larger than the ESP system size. However, to achieve the same precision $\epsilon$ for a simulation time $T$, the ESP system would need to exponentially increase in system size. Thus, overall, the Hermite system size is still exponentially smaller than the ESP system size. It is important to note that the system size's dependence on $T$ exists only in the collisionless case. Upon adding collisions, the number of Hermite moments required to capture the fine structures is dictated by $M \gtrsim \omega/\nu$, which is independent of $T$.

\section{\label{sec:discon}Discussion and Conclusion}

In this paper we have demonstrated two main results. First, using a Hermite representation of velocity space, we can obtain a system that is exponentially smaller than the one obtained via finite-difference discretization, as proposed in the work of ESP~\cite{engel2019quantum}, for the same error $\epsilon$. This implies that a classical implementation of the Hermite approach will have similar performance to that of ESP's quantum algorithm. 
Second, a quantum algorithm for the Hermite formulation can yield a quadratic speedup compared to classical algorithms that solve the same system of equations. 
An exponential speedup, however, does not seem possible with currently known methods due to the large norm of the matrices involved. Table \ref{tab:results} summarizes the complexities of the algorithms discussed in this paper. 
\begin{table}[h!]
    \centering
    \begin{tabular}{c|c|c}
        \hline
        & \multicolumn{2}{c}{Gate Complexity} \\
        Representation & Classical & Quantum\\
        \hline \hline
        ESP~\cite{engel2019quantum} & $\mathcal{O} (N_v/\epsilon^\theta)$ & $\mathcal{O}(\mathrm{polylog} (N_v) /\epsilon)$ \\
        Hermite & $\mathcal{O} (N /\epsilon^\theta)$ & $\mathcal{O} (\sqrt{N} \log (N)/\epsilon)$ \\
        \hline
    \end{tabular}
    \caption{Gate complexities of the algorithms to estimate Landau damping discussed in this paper. Here $N$ is the Hermite system size, $N_v$ is the ESP system size, $\epsilon$ is the absolute error in amplitude estimation, and $\theta \leq 1$ is the order of the classical ODE solver, with smaller values corresponding to higher-order solvers. The simulation time $T$ is a constant and is omitted from the complexity analysis.}
    \label{tab:results}
\end{table}

The problem analyzed in this paper is somewhat particular in that system size is conflated with the computation error.
This is because the resolution required in velocity (or Hermite) space is strictly a function only of the error one wishes to achieve in the computation of the Landau damping rate (physically, running the computation for longer times so that a longer-in-time decay stage is obtained leads to more phase-mixing and thus finer-scale structure in velocity space --- or, correspondingly, structure at higher Hermite moments).
In more general problems in plasma physics, there are minimal requirements imposed on velocity- and position-space grid sizes, set by the need to resolve specific physics processes (for example, one typically wishes to simulate a system of a given size, $L$, but is forced to resolve kinetic-scale physics happening at scales below, say, the ion Larmor radius, $\rho_i$. Frequently, $L/\rho_i\gg1$, implying, therefore, a very large number of grid points in position space before one can even consider the error convergence with respect to system size. 
A comparable situation in velocity space is one where there is a super-thermal particle population, in addition to a Maxwellian bulk).
Therefore, the scaling with system size of quantum algorithms for plasma problems is of intrinsic interest.

Our results also highlight the challenge of applying currently existing quantum algorithms to real-world problems. In applying quantum algorithms for Hamiltonian simulation or differential equation solvers to the Vlasov equation, we encounter matrix norms that scale as the square-root of the system size. In addition, extracting classical information such as the Landau damping parameter increases the complexity from $\mathcal{O}\left(\log(1/\epsilon)\right)$ to $\mathcal{O}(1/\epsilon)$.

We also note that it is possible to formulate the linear Vlasov equation as an eigenvalue problem and use quantum phase estimation \cite{kitaev1995quantum,nielsen2002quantum} and quantum eigenvalue solvers \cite{abrams1999quantum,jaksch2003eigenvector,wang2010measurement,daskin2014universal,shao2022computing} to determine the eigenvalue spectrum for the collisionless and collisional systems, respectively. However, in the collisionless case, finding the eigenvalue spectrum requires $\mathcal{O}(N)$ repetitions, and in the collisional case, the complexity of such algorithms depends on the condition number of the diagonalizing matrix of the system, which, in this case, grows exponentially with $N$. Furthermore, Landau damping cannot be captured with an eigenvalue formulation, as it inherently requires an initial-value problem approach.

Possible extensions of this work include generalizing the system to higher dimensions, as well as extending the quantum algorithm to the fully nonlinear Vlasov equation. For the latter, in the regime of weak nonlinearity, it is possible that quantum approaches such as those proposed by Liu \textit{et al.}~\cite{liu2021efficient} could lead to a speedup. 
Whether a quantum speedup can be obtained in a strongly nonlinear regime remains an open and important problem.
\newline
\paragraph*{Acknowledgements.}
AA thanks Alexander Engel and Scott Parker for providing great insight on their work and for the useful feedback on this paper, as well as Muni Zhou and Noah Mandell for helpful discussions on plasma physics and quantum algorithms. AA acknowledges support from NSERC PGS-D. The authors acknowledge support from the US Department of Energy grant no. DE-SC0020264.

\bibliography{apssamp}

\providecommand{\noopsort}[1]{}\providecommand{\singleletter}[1]{\#1}%
\begin{thebibliography}{60}%
\makeatletter
\providecommand \@ifxundefined [1]{%
 \@ifx{#1\undefined}
}%
\providecommand \@ifnum [1]{%
 \ifnum #1\expandafter \@firstoftwo
 \else \expandafter \@secondoftwo
 \fi
}%
\providecommand \@ifx [1]{%
 \ifx #1\expandafter \@firstoftwo
 \else \expandafter \@secondoftwo
 \fi
}%
\providecommand \natexlab [1]{#1}%
\providecommand \enquote  [1]{``#1''}%
\providecommand \bibnamefont  [1]{#1}%
\providecommand \bibfnamefont [1]{#1}%
\providecommand \citenamefont [1]{#1}%
\providecommand \href@noop [0]{\@secondoftwo}%
\providecommand \href [0]{\begingroup \@sanitize@url \@href}%
\providecommand \@href[1]{\@@startlink{#1}\@@href}%
\providecommand \@@href[1]{\endgroup#1\@@endlink}%
\providecommand \@sanitize@url [0]{\catcode `\\12\catcode `\$12\catcode
  `\&12\catcode `\#12\catcode `\^12\catcode `\_12\catcode `\%12\relax}%
\providecommand \@@startlink[1]{}%
\providecommand \@@endlink[0]{}%
\providecommand \url  [0]{\begingroup\@sanitize@url \@url }%
\providecommand \@url [1]{\endgroup\@href {#1}{\urlprefix }}%
\providecommand \urlprefix  [0]{URL }%
\providecommand \Eprint [0]{\href }%
\providecommand \doibase [0]{https://doi.org/}%
\providecommand \selectlanguage [0]{\@gobble}%
\providecommand \bibinfo  [0]{\@secondoftwo}%
\providecommand \bibfield  [0]{\@secondoftwo}%
\providecommand \translation [1]{[#1]}%
\providecommand \BibitemOpen [0]{}%
\providecommand \bibitemStop [0]{}%
\providecommand \bibitemNoStop [0]{.\EOS\space}%
\providecommand \EOS [0]{\spacefactor3000\relax}%
\providecommand \BibitemShut  [1]{\csname bibitem#1\endcsname}%
\let\auto@bib@innerbib\@empty
\bibitem [{\citenamefont {Grover}(1996)}]{grover1996fast}%
  \BibitemOpen
  \bibfield  {author} {\bibinfo {author} {\bibfnamefont {L.~K.}\ \bibnamefont
  {Grover}},\ }\bibfield  {title} {\bibinfo {title} {A fast quantum mechanical
  algorithm for database search},\ }in\ \href
  {https://doi.org/10.48550/arXiv.quant-ph/9605043} {\emph {\bibinfo
  {booktitle} {Proceedings of the twenty-eighth annual ACM symposium on Theory
  of computing}}}\ (\bibinfo {year} {1996})\ pp.\ \bibinfo {pages}
  {212--219}\BibitemShut {NoStop}%
\bibitem [{\citenamefont {Lloyd}(1996)}]{lloyd1996universal}%
  \BibitemOpen
  \bibfield  {author} {\bibinfo {author} {\bibfnamefont {S.}~\bibnamefont
  {Lloyd}},\ }\bibfield  {title} {\bibinfo {title} {Universal quantum
  simulators},\ }\href {https://doi.org/10.1126/science.273.5278.1073}
  {\bibfield  {journal} {\bibinfo  {journal} {Science}\ }\textbf {\bibinfo
  {volume} {273}},\ \bibinfo {pages} {1073} (\bibinfo {year}
  {1996})}\BibitemShut {NoStop}%
\bibitem [{\citenamefont {Berry}\ \emph {et~al.}(2007)\citenamefont {Berry},
  \citenamefont {Ahokas}, \citenamefont {Cleve},\ and\ \citenamefont
  {Sanders}}]{berry2007efficient}%
  \BibitemOpen
  \bibfield  {author} {\bibinfo {author} {\bibfnamefont {D.~W.}\ \bibnamefont
  {Berry}}, \bibinfo {author} {\bibfnamefont {G.}~\bibnamefont {Ahokas}},
  \bibinfo {author} {\bibfnamefont {R.}~\bibnamefont {Cleve}},\ and\ \bibinfo
  {author} {\bibfnamefont {B.~C.}\ \bibnamefont {Sanders}},\ }\bibfield
  {title} {\bibinfo {title} {Efficient quantum algorithms for simulating sparse
  {H}amiltonians},\ }\href {https://doi.org/10.1007/s00220-006-0150-x}
  {\bibfield  {journal} {\bibinfo  {journal} {Communications in Mathematical
  Physics}\ }\textbf {\bibinfo {volume} {270}},\ \bibinfo {pages} {359}
  (\bibinfo {year} {2007})}\BibitemShut {NoStop}%
\bibitem [{\citenamefont {Childs}\ and\ \citenamefont
  {Wiebe}(2012)}]{childs2012hamiltonian}%
  \BibitemOpen
  \bibfield  {author} {\bibinfo {author} {\bibfnamefont {A.~M.}\ \bibnamefont
  {Childs}}\ and\ \bibinfo {author} {\bibfnamefont {N.}~\bibnamefont {Wiebe}},\
  }\bibfield  {title} {\bibinfo {title} {Hamiltonian simulation using linear
  combinations of unitary operations},\ }\href
  {https://doi.org/10.48550/arXiv.1202.5822} {\bibfield  {journal} {\bibinfo
  {journal} {arXiv preprint arXiv:1202.5822}\ } (\bibinfo {year}
  {2012})}\BibitemShut {NoStop}%
\bibitem [{\citenamefont {Berry}\ \emph {et~al.}(2014)\citenamefont {Berry},
  \citenamefont {Childs}, \citenamefont {Cleve}, \citenamefont {Kothari},\ and\
  \citenamefont {Somma}}]{berry2014exponential}%
  \BibitemOpen
  \bibfield  {author} {\bibinfo {author} {\bibfnamefont {D.~W.}\ \bibnamefont
  {Berry}}, \bibinfo {author} {\bibfnamefont {A.~M.}\ \bibnamefont {Childs}},
  \bibinfo {author} {\bibfnamefont {R.}~\bibnamefont {Cleve}}, \bibinfo
  {author} {\bibfnamefont {R.}~\bibnamefont {Kothari}},\ and\ \bibinfo {author}
  {\bibfnamefont {R.~D.}\ \bibnamefont {Somma}},\ }\bibfield  {title} {\bibinfo
  {title} {Exponential improvement in precision for simulating sparse
  {H}amiltonians},\ }in\ \href {https://doi.org/10.1145/2591796.2591854} {\emph
  {\bibinfo {booktitle} {Proceedings of the forty-sixth annual ACM symposium on
  Theory of computing}}}\ (\bibinfo {year} {2014})\ pp.\ \bibinfo {pages}
  {283--292}\BibitemShut {NoStop}%
\bibitem [{\citenamefont {Berry}\ \emph {et~al.}(2015)\citenamefont {Berry},
  \citenamefont {Childs}, \citenamefont {Cleve}, \citenamefont {Kothari},\ and\
  \citenamefont {Somma}}]{berry2015simulating}%
  \BibitemOpen
  \bibfield  {author} {\bibinfo {author} {\bibfnamefont {D.~W.}\ \bibnamefont
  {Berry}}, \bibinfo {author} {\bibfnamefont {A.~M.}\ \bibnamefont {Childs}},
  \bibinfo {author} {\bibfnamefont {R.}~\bibnamefont {Cleve}}, \bibinfo
  {author} {\bibfnamefont {R.}~\bibnamefont {Kothari}},\ and\ \bibinfo {author}
  {\bibfnamefont {R.~D.}\ \bibnamefont {Somma}},\ }\bibfield  {title} {\bibinfo
  {title} {Simulating {H}amiltonian dynamics with a truncated {T}aylor
  series},\ }\href {https://doi.org/10.1103/PhysRevLett.114.090502} {\bibfield
  {journal} {\bibinfo  {journal} {Physical Review Letters}\ }\textbf {\bibinfo
  {volume} {114}},\ \bibinfo {pages} {090502} (\bibinfo {year}
  {2015})}\BibitemShut {NoStop}%
\bibitem [{\citenamefont {Low}\ and\ \citenamefont
  {Chuang}(2017)}]{low2017optimal}%
  \BibitemOpen
  \bibfield  {author} {\bibinfo {author} {\bibfnamefont {G.~H.}\ \bibnamefont
  {Low}}\ and\ \bibinfo {author} {\bibfnamefont {I.~L.}\ \bibnamefont
  {Chuang}},\ }\bibfield  {title} {\bibinfo {title} {Optimal {H}amiltonian
  simulation by quantum signal processing},\ }\href
  {https://doi.org/10.1103/PhysRevLett.118.010501} {\bibfield  {journal}
  {\bibinfo  {journal} {Physical Review Letters}\ }\textbf {\bibinfo {volume}
  {118}},\ \bibinfo {pages} {010501} (\bibinfo {year} {2017})}\BibitemShut
  {NoStop}%
\bibitem [{\citenamefont {Low}\ and\ \citenamefont
  {Chuang}(2019)}]{low2019hamiltonian}%
  \BibitemOpen
  \bibfield  {author} {\bibinfo {author} {\bibfnamefont {G.~H.}\ \bibnamefont
  {Low}}\ and\ \bibinfo {author} {\bibfnamefont {I.~L.}\ \bibnamefont
  {Chuang}},\ }\bibfield  {title} {\bibinfo {title} {Hamiltonian simulation by
  qubitization},\ }\href {https://doi.org/10.22331/q-2019-07-12-163} {\bibfield
   {journal} {\bibinfo  {journal} {Quantum}\ }\textbf {\bibinfo {volume} {3}},\
  \bibinfo {pages} {163} (\bibinfo {year} {2019})}\BibitemShut {NoStop}%
\bibitem [{\citenamefont {Shor}(1999)}]{shor1999polynomial}%
  \BibitemOpen
  \bibfield  {author} {\bibinfo {author} {\bibfnamefont {P.~W.}\ \bibnamefont
  {Shor}},\ }\bibfield  {title} {\bibinfo {title} {Polynomial-time algorithms
  for prime factorization and discrete logarithms on a quantum computer},\
  }\href {https://doi.org/10.1137/S0036144598347011} {\bibfield  {journal}
  {\bibinfo  {journal} {SIAM review}\ }\textbf {\bibinfo {volume} {41}},\
  \bibinfo {pages} {303} (\bibinfo {year} {1999})}\BibitemShut {NoStop}%
\bibitem [{\citenamefont {Jordan}\ \emph {et~al.}(2012)\citenamefont {Jordan},
  \citenamefont {Lee},\ and\ \citenamefont {Preskill}}]{JLP12}%
  \BibitemOpen
  \bibfield  {author} {\bibinfo {author} {\bibfnamefont {S.~P.}\ \bibnamefont
  {Jordan}}, \bibinfo {author} {\bibfnamefont {K.~S.~M.}\ \bibnamefont {Lee}},\
  and\ \bibinfo {author} {\bibfnamefont {J.}~\bibnamefont {Preskill}},\
  }\bibfield  {title} {\bibinfo {title} {Quantum algorithms for quantum field
  theories},\ }\href {https://doi.org/10.1126/science.1217069} {\bibfield
  {journal} {\bibinfo  {journal} {Science}\ }\textbf {\bibinfo {volume}
  {336}},\ \bibinfo {pages} {1130} (\bibinfo {year} {2012})}\BibitemShut
  {NoStop}%
\bibitem [{\citenamefont {Jordan}\ \emph {et~al.}(2014)\citenamefont {Jordan},
  \citenamefont {Lee},\ and\ \citenamefont {Preskill}}]{fermionic_qft}%
  \BibitemOpen
  \bibfield  {author} {\bibinfo {author} {\bibfnamefont {S.~P.}\ \bibnamefont
  {Jordan}}, \bibinfo {author} {\bibfnamefont {K.~S.}\ \bibnamefont {Lee}},\
  and\ \bibinfo {author} {\bibfnamefont {J.}~\bibnamefont {Preskill}},\
  }\bibfield  {title} {\bibinfo {title} {Quantum algorithms for fermionic
  quantum field theories},\ }\href {https://arxiv.org/abs/1404.7115} {\bibfield
   {journal} {\bibinfo  {journal} {arXiv preprint arXiv:1404.7115}\ } (\bibinfo
  {year} {2014})}\BibitemShut {NoStop}%
\bibitem [{\citenamefont {Osborne}\ and\ \citenamefont
  {Stottmeister}(2021)}]{cft}%
  \BibitemOpen
  \bibfield  {author} {\bibinfo {author} {\bibfnamefont {T.~J.}\ \bibnamefont
  {Osborne}}\ and\ \bibinfo {author} {\bibfnamefont {A.}~\bibnamefont
  {Stottmeister}},\ }\bibfield  {title} {\bibinfo {title} {Quantum simulation
  of conformal field theory},\ }\href {https://arxiv.org/abs/2109.14214}
  {\bibfield  {journal} {\bibinfo  {journal} {arXiv preprint arXiv:2109.14214}\
  } (\bibinfo {year} {2021})}\BibitemShut {NoStop}%
\bibitem [{\citenamefont {Harrow}\ \emph {et~al.}(2009)\citenamefont {Harrow},
  \citenamefont {Hassidim},\ and\ \citenamefont {Lloyd}}]{harrow2009quantum}%
  \BibitemOpen
  \bibfield  {author} {\bibinfo {author} {\bibfnamefont {A.~W.}\ \bibnamefont
  {Harrow}}, \bibinfo {author} {\bibfnamefont {A.}~\bibnamefont {Hassidim}},\
  and\ \bibinfo {author} {\bibfnamefont {S.}~\bibnamefont {Lloyd}},\ }\bibfield
   {title} {\bibinfo {title} {Quantum algorithm for linear systems of
  equations},\ }\href {https://doi.org/10.1103/PhysRevLett.103.150502}
  {\bibfield  {journal} {\bibinfo  {journal} {Physical Review Letters}\
  }\textbf {\bibinfo {volume} {103}},\ \bibinfo {pages} {150502} (\bibinfo
  {year} {2009})}\BibitemShut {NoStop}%
\bibitem [{\citenamefont {Childs}\ \emph {et~al.}(2017)\citenamefont {Childs},
  \citenamefont {Kothari},\ and\ \citenamefont {Somma}}]{childs2017quantum}%
  \BibitemOpen
  \bibfield  {author} {\bibinfo {author} {\bibfnamefont {A.~M.}\ \bibnamefont
  {Childs}}, \bibinfo {author} {\bibfnamefont {R.}~\bibnamefont {Kothari}},\
  and\ \bibinfo {author} {\bibfnamefont {R.~D.}\ \bibnamefont {Somma}},\
  }\bibfield  {title} {\bibinfo {title} {Quantum algorithm for systems of
  linear equations with exponentially improved dependence on precision},\
  }\href {https://doi.org/10.1137/16M1087072} {\bibfield  {journal} {\bibinfo
  {journal} {SIAM Journal on Computing}\ }\textbf {\bibinfo {volume} {46}},\
  \bibinfo {pages} {1920} (\bibinfo {year} {2017})}\BibitemShut {NoStop}%
\bibitem [{\citenamefont {Berry}(2014)}]{berry2014high}%
  \BibitemOpen
  \bibfield  {author} {\bibinfo {author} {\bibfnamefont {D.~W.}\ \bibnamefont
  {Berry}},\ }\bibfield  {title} {\bibinfo {title} {High-order quantum
  algorithm for solving linear differential equations},\ }\href
  {https://doi.org/10.1088/1751-8113/47/10/105301} {\bibfield  {journal}
  {\bibinfo  {journal} {Journal of Physics A: Mathematical and Theoretical}\
  }\textbf {\bibinfo {volume} {47}},\ \bibinfo {pages} {105301} (\bibinfo
  {year} {2014})}\BibitemShut {NoStop}%
\bibitem [{\citenamefont {Berry}\ \emph {et~al.}(2017)\citenamefont {Berry},
  \citenamefont {Childs}, \citenamefont {Ostrander},\ and\ \citenamefont
  {Wang}}]{BCOW}%
  \BibitemOpen
  \bibfield  {author} {\bibinfo {author} {\bibfnamefont {D.~W.}\ \bibnamefont
  {Berry}}, \bibinfo {author} {\bibfnamefont {A.~M.}\ \bibnamefont {Childs}},
  \bibinfo {author} {\bibfnamefont {A.}~\bibnamefont {Ostrander}},\ and\
  \bibinfo {author} {\bibfnamefont {G.}~\bibnamefont {Wang}},\ }\bibfield
  {title} {\bibinfo {title} {Quantum algorithm for linear differential
  equations with exponentially improved dependence on precision},\ }\href
  {https://doi.org/10.1007/s00220-017-3002-y} {\bibfield  {journal} {\bibinfo
  {journal} {Communications in Mathematical Physics}\ }\textbf {\bibinfo
  {volume} {356}},\ \bibinfo {pages} {1057} (\bibinfo {year}
  {2017})}\BibitemShut {NoStop}%
\bibitem [{\citenamefont {Childs}\ and\ \citenamefont
  {Liu}(2020)}]{childs2020}%
  \BibitemOpen
  \bibfield  {author} {\bibinfo {author} {\bibfnamefont {A.~M.}\ \bibnamefont
  {Childs}}\ and\ \bibinfo {author} {\bibfnamefont {J.-P.}\ \bibnamefont
  {Liu}},\ }\bibfield  {title} {\bibinfo {title} {Quantum spectral methods for
  differential equations},\ }\href {https://doi.org/10.1007/s00220-020-03699-z}
  {\bibfield  {journal} {\bibinfo  {journal} {Communications in Mathematical
  Physics}\ }\textbf {\bibinfo {volume} {375}},\ \bibinfo {pages} {1427–1457}
  (\bibinfo {year} {2020})}\BibitemShut {NoStop}%
\bibitem [{\citenamefont {Krovi}(2023)}]{krovi2022improved}%
  \BibitemOpen
  \bibfield  {author} {\bibinfo {author} {\bibfnamefont {H.}~\bibnamefont
  {Krovi}},\ }\bibfield  {title} {\bibinfo {title} {Improved quantum algorithms
  for linear and nonlinear differential equations},\ }\href
  {https://doi.org/10.22331/q-2023-02-02-913} {\bibfield  {journal} {\bibinfo
  {journal} {{Quantum}}\ }\textbf {\bibinfo {volume} {7}},\ \bibinfo {pages}
  {913} (\bibinfo {year} {2023})}\BibitemShut {NoStop}%
\bibitem [{\citenamefont {Leyton}\ and\ \citenamefont
  {Osborne}(2008)}]{leyton2008quantum}%
  \BibitemOpen
  \bibfield  {author} {\bibinfo {author} {\bibfnamefont {S.~K.}\ \bibnamefont
  {Leyton}}\ and\ \bibinfo {author} {\bibfnamefont {T.~J.}\ \bibnamefont
  {Osborne}},\ }\bibfield  {title} {\bibinfo {title} {A quantum algorithm to
  solve nonlinear differential equations},\ }\href
  {https://doi.org/10.48550/arXiv.0812.4423} {\bibfield  {journal} {\bibinfo
  {journal} {arXiv preprint arXiv:0812.4423}\ } (\bibinfo {year}
  {2008})}\BibitemShut {NoStop}%
\bibitem [{\citenamefont {Lloyd}\ \emph {et~al.}(2020)\citenamefont {Lloyd},
  \citenamefont {De~Palma}, \citenamefont {Gokler}, \citenamefont {Kiani},
  \citenamefont {Liu}, \citenamefont {Marvian}, \citenamefont {Tennie},\ and\
  \citenamefont {Palmer}}]{lloyd2020quantum}%
  \BibitemOpen
  \bibfield  {author} {\bibinfo {author} {\bibfnamefont {S.}~\bibnamefont
  {Lloyd}}, \bibinfo {author} {\bibfnamefont {G.}~\bibnamefont {De~Palma}},
  \bibinfo {author} {\bibfnamefont {C.}~\bibnamefont {Gokler}}, \bibinfo
  {author} {\bibfnamefont {B.}~\bibnamefont {Kiani}}, \bibinfo {author}
  {\bibfnamefont {Z.-W.}\ \bibnamefont {Liu}}, \bibinfo {author} {\bibfnamefont
  {M.}~\bibnamefont {Marvian}}, \bibinfo {author} {\bibfnamefont
  {F.}~\bibnamefont {Tennie}},\ and\ \bibinfo {author} {\bibfnamefont
  {T.}~\bibnamefont {Palmer}},\ }\bibfield  {title} {\bibinfo {title} {Quantum
  algorithm for nonlinear differential equations},\ }\href
  {https://doi.org/10.48550/arXiv.2011.06571} {\bibfield  {journal} {\bibinfo
  {journal} {arXiv preprint arXiv:2011.06571}\ } (\bibinfo {year}
  {2020})}\BibitemShut {NoStop}%
\bibitem [{\citenamefont {Liu}\ \emph {et~al.}(2021)\citenamefont {Liu},
  \citenamefont {Kolden}, \citenamefont {Krovi}, \citenamefont {Loureiro},
  \citenamefont {Trivisa},\ and\ \citenamefont {Childs}}]{liu2021efficient}%
  \BibitemOpen
  \bibfield  {author} {\bibinfo {author} {\bibfnamefont {J.-P.}\ \bibnamefont
  {Liu}}, \bibinfo {author} {\bibfnamefont {H.~{\O}.}\ \bibnamefont {Kolden}},
  \bibinfo {author} {\bibfnamefont {H.~K.}\ \bibnamefont {Krovi}}, \bibinfo
  {author} {\bibfnamefont {N.~F.}\ \bibnamefont {Loureiro}}, \bibinfo {author}
  {\bibfnamefont {K.}~\bibnamefont {Trivisa}},\ and\ \bibinfo {author}
  {\bibfnamefont {A.~M.}\ \bibnamefont {Childs}},\ }\bibfield  {title}
  {\bibinfo {title} {Efficient quantum algorithm for dissipative nonlinear
  differential equations},\ }\href {https://doi.org/10.1073/pnas.2026805118}
  {\bibfield  {journal} {\bibinfo  {journal} {Proceedings of the National
  Academy of Sciences}\ }\textbf {\bibinfo {volume} {118}} (\bibinfo {year}
  {2021})}\BibitemShut {NoStop}%
\bibitem [{\citenamefont {Xue}\ \emph {et~al.}(2021)\citenamefont {Xue},
  \citenamefont {Yu-Chun},\ and\ \citenamefont {Guo}}]{xue2021quantum}%
  \BibitemOpen
  \bibfield  {author} {\bibinfo {author} {\bibfnamefont {C.}~\bibnamefont
  {Xue}}, \bibinfo {author} {\bibfnamefont {W.}~\bibnamefont {Yu-Chun}},\ and\
  \bibinfo {author} {\bibfnamefont {G.}~\bibnamefont {Guo}},\ }\bibfield
  {title} {\bibinfo {title} {Quantum homotopy perturbation method for nonlinear
  dissipative ordinary differential equations},\ }\href
  {https://doi.org/10.1088/1367-2630/ac3eff} {\bibfield  {journal} {\bibinfo
  {journal} {New Journal of Physics}\ } (\bibinfo {year} {2021})}\BibitemShut
  {NoStop}%
\bibitem [{\citenamefont {Kyriienko}\ \emph {et~al.}(2021)\citenamefont
  {Kyriienko}, \citenamefont {Paine},\ and\ \citenamefont
  {Elfving}}]{kyriienko2021solving}%
  \BibitemOpen
  \bibfield  {author} {\bibinfo {author} {\bibfnamefont {O.}~\bibnamefont
  {Kyriienko}}, \bibinfo {author} {\bibfnamefont {A.~E.}\ \bibnamefont
  {Paine}},\ and\ \bibinfo {author} {\bibfnamefont {V.~E.}\ \bibnamefont
  {Elfving}},\ }\bibfield  {title} {\bibinfo {title} {Solving nonlinear
  differential equations with differentiable quantum circuits},\ }\href
  {https://doi.org/10.1103/PhysRevA.103.052416} {\bibfield  {journal} {\bibinfo
   {journal} {Physical Review A}\ }\textbf {\bibinfo {volume} {103}},\ \bibinfo
  {pages} {052416} (\bibinfo {year} {2021})}\BibitemShut {NoStop}%
\bibitem [{\citenamefont {Cao}\ \emph {et~al.}(2013)\citenamefont {Cao},
  \citenamefont {Papageorgiou}, \citenamefont {Petras}, \citenamefont {Traub},\
  and\ \citenamefont {Kais}}]{cao2013quantum}%
  \BibitemOpen
  \bibfield  {author} {\bibinfo {author} {\bibfnamefont {Y.}~\bibnamefont
  {Cao}}, \bibinfo {author} {\bibfnamefont {A.}~\bibnamefont {Papageorgiou}},
  \bibinfo {author} {\bibfnamefont {I.}~\bibnamefont {Petras}}, \bibinfo
  {author} {\bibfnamefont {J.}~\bibnamefont {Traub}},\ and\ \bibinfo {author}
  {\bibfnamefont {S.}~\bibnamefont {Kais}},\ }\bibfield  {title} {\bibinfo
  {title} {Quantum algorithm and circuit design solving the {P}oisson
  equation},\ }\href {https://doi.org/10.1088/1367-2630/15/1/013021} {\bibfield
   {journal} {\bibinfo  {journal} {New Journal of Physics}\ }\textbf {\bibinfo
  {volume} {15}},\ \bibinfo {pages} {013021} (\bibinfo {year}
  {2013})}\BibitemShut {NoStop}%
\bibitem [{\citenamefont {Costa}\ \emph {et~al.}(2019)\citenamefont {Costa},
  \citenamefont {Jordan},\ and\ \citenamefont {Ostrander}}]{costa2019quantum}%
  \BibitemOpen
  \bibfield  {author} {\bibinfo {author} {\bibfnamefont {P.~C.}\ \bibnamefont
  {Costa}}, \bibinfo {author} {\bibfnamefont {S.}~\bibnamefont {Jordan}},\ and\
  \bibinfo {author} {\bibfnamefont {A.}~\bibnamefont {Ostrander}},\ }\bibfield
  {title} {\bibinfo {title} {Quantum algorithm for simulating the wave
  equation},\ }\href {https://doi.org/10.1103/PhysRevA.99.012323} {\bibfield
  {journal} {\bibinfo  {journal} {Physical Review A}\ }\textbf {\bibinfo
  {volume} {99}},\ \bibinfo {pages} {012323} (\bibinfo {year}
  {2019})}\BibitemShut {NoStop}%
\bibitem [{\citenamefont {Dodin}\ and\ \citenamefont
  {Startsev}(2021)}]{dodin2021applications}%
  \BibitemOpen
  \bibfield  {author} {\bibinfo {author} {\bibfnamefont {I.~Y.}\ \bibnamefont
  {Dodin}}\ and\ \bibinfo {author} {\bibfnamefont {E.~A.}\ \bibnamefont
  {Startsev}},\ }\bibfield  {title} {\bibinfo {title} {On applications of
  quantum computing to plasma simulations},\ }\href
  {https://doi.org/10.1063/5.0056974} {\bibfield  {journal} {\bibinfo
  {journal} {Physics of Plasmas}\ }\textbf {\bibinfo {volume} {28}},\ \bibinfo
  {pages} {092101} (\bibinfo {year} {2021})}\BibitemShut {NoStop}%
\bibitem [{\citenamefont {Zylberman}\ \emph {et~al.}(2022)\citenamefont
  {Zylberman}, \citenamefont {Di~Molfetta}, \citenamefont {Brachet},
  \citenamefont {Loureiro},\ and\ \citenamefont
  {Debbasch}}]{zylberman2022hybrid}%
  \BibitemOpen
  \bibfield  {author} {\bibinfo {author} {\bibfnamefont {J.}~\bibnamefont
  {Zylberman}}, \bibinfo {author} {\bibfnamefont {G.}~\bibnamefont
  {Di~Molfetta}}, \bibinfo {author} {\bibfnamefont {M.}~\bibnamefont
  {Brachet}}, \bibinfo {author} {\bibfnamefont {N.~F.}\ \bibnamefont
  {Loureiro}},\ and\ \bibinfo {author} {\bibfnamefont {F.}~\bibnamefont
  {Debbasch}},\ }\bibfield  {title} {\bibinfo {title} {Hybrid quantum-classical
  algorithm for hydrodynamics},\ }\href
  {https://doi.org/10.48550/arXiv.2202.00918} {\bibfield  {journal} {\bibinfo
  {journal} {arXiv preprint arXiv:2202.00918}\ } (\bibinfo {year}
  {2022})}\BibitemShut {NoStop}%
\bibitem [{\citenamefont {Engel}\ \emph {et~al.}(2021)\citenamefont {Engel},
  \citenamefont {Smith},\ and\ \citenamefont {Parker}}]{engel2021linear}%
  \BibitemOpen
  \bibfield  {author} {\bibinfo {author} {\bibfnamefont {A.}~\bibnamefont
  {Engel}}, \bibinfo {author} {\bibfnamefont {G.}~\bibnamefont {Smith}},\ and\
  \bibinfo {author} {\bibfnamefont {S.~E.}\ \bibnamefont {Parker}},\ }\bibfield
   {title} {\bibinfo {title} {Linear embedding of nonlinear dynamical systems
  and prospects for efficient quantum algorithms},\ }\href
  {https://aip.scitation.org/doi/full/10.1063/5.0040313?casa_token=b2tHl4_sjuEAAAAA\%3AKGnla2-3ZAKDptmYGDAVaD3hWpQ0m5pBm5v6Rt5Hobs1CkViYp-8FOjlJMF4sqOI7wvqvD_EVI0}
  {\bibfield  {journal} {\bibinfo  {journal} {Physics of Plasmas}\ }\textbf
  {\bibinfo {volume} {28}},\ \bibinfo {pages} {062305} (\bibinfo {year}
  {2021})}\BibitemShut {NoStop}%
\bibitem [{\citenamefont {Joseph}(2020)}]{joseph2020koopman}%
  \BibitemOpen
  \bibfield  {author} {\bibinfo {author} {\bibfnamefont {I.}~\bibnamefont
  {Joseph}},\ }\bibfield  {title} {\bibinfo {title} {Koopman--von {N}eumann
  approach to quantum simulation of nonlinear classical dynamics},\ }\href
  {https://journals.aps.org/prresearch/abstract/10.1103/PhysRevResearch.2.043102}
  {\bibfield  {journal} {\bibinfo  {journal} {Physical Review Research}\
  }\textbf {\bibinfo {volume} {2}},\ \bibinfo {pages} {043102} (\bibinfo {year}
  {2020})}\BibitemShut {NoStop}%
\bibitem [{\citenamefont {Joseph}\ \emph {et~al.}(2023)\citenamefont {Joseph},
  \citenamefont {Shi}, \citenamefont {Porter}, \citenamefont {Castelli},
  \citenamefont {Geyko}, \citenamefont {Graziani}, \citenamefont {Libby},\ and\
  \citenamefont {DuBois}}]{joseph2023quantum}%
  \BibitemOpen
  \bibfield  {author} {\bibinfo {author} {\bibfnamefont {I.}~\bibnamefont
  {Joseph}}, \bibinfo {author} {\bibfnamefont {Y.}~\bibnamefont {Shi}},
  \bibinfo {author} {\bibfnamefont {M.}~\bibnamefont {Porter}}, \bibinfo
  {author} {\bibfnamefont {A.}~\bibnamefont {Castelli}}, \bibinfo {author}
  {\bibfnamefont {V.}~\bibnamefont {Geyko}}, \bibinfo {author} {\bibfnamefont
  {F.}~\bibnamefont {Graziani}}, \bibinfo {author} {\bibfnamefont
  {S.}~\bibnamefont {Libby}},\ and\ \bibinfo {author} {\bibfnamefont
  {J.}~\bibnamefont {DuBois}},\ }\bibfield  {title} {\bibinfo {title} {Quantum
  computing for fusion energy science applications},\ }\href
  {https://doi.org/10.1063/5.0123765} {\bibfield  {journal} {\bibinfo
  {journal} {Physics of Plasmas}\ }\textbf {\bibinfo {volume} {30}},\ \bibinfo
  {pages} {010501} (\bibinfo {year} {2023})}\BibitemShut {NoStop}%
\bibitem [{\citenamefont {Shi}\ \emph {et~al.}(2021)\citenamefont {Shi},
  \citenamefont {Castelli}, \citenamefont {Wu}, \citenamefont {Joseph},
  \citenamefont {Geyko}, \citenamefont {Graziani}, \citenamefont {Libby},
  \citenamefont {Parker}, \citenamefont {Rosen}, \citenamefont {Martinez} \emph
  {et~al.}}]{shi2021simulating}%
  \BibitemOpen
  \bibfield  {author} {\bibinfo {author} {\bibfnamefont {Y.}~\bibnamefont
  {Shi}}, \bibinfo {author} {\bibfnamefont {A.~R.}\ \bibnamefont {Castelli}},
  \bibinfo {author} {\bibfnamefont {X.}~\bibnamefont {Wu}}, \bibinfo {author}
  {\bibfnamefont {I.}~\bibnamefont {Joseph}}, \bibinfo {author} {\bibfnamefont
  {V.}~\bibnamefont {Geyko}}, \bibinfo {author} {\bibfnamefont {F.~R.}\
  \bibnamefont {Graziani}}, \bibinfo {author} {\bibfnamefont {S.~B.}\
  \bibnamefont {Libby}}, \bibinfo {author} {\bibfnamefont {J.~B.}\ \bibnamefont
  {Parker}}, \bibinfo {author} {\bibfnamefont {Y.~J.}\ \bibnamefont {Rosen}},
  \bibinfo {author} {\bibfnamefont {L.~A.}\ \bibnamefont {Martinez}}, \emph
  {et~al.},\ }\bibfield  {title} {\bibinfo {title} {Simulating non-native cubic
  interactions on noisy quantum machines},\ }\href
  {https://doi.org/10.1103/PhysRevA.103.062608} {\bibfield  {journal} {\bibinfo
   {journal} {Physical Review A}\ }\textbf {\bibinfo {volume} {103}},\ \bibinfo
  {pages} {062608} (\bibinfo {year} {2021})}\BibitemShut {NoStop}%
\bibitem [{\citenamefont {Novikau}\ \emph {et~al.}(2022)\citenamefont
  {Novikau}, \citenamefont {Startsev},\ and\ \citenamefont
  {Dodin}}]{novikau2021quantum}%
  \BibitemOpen
  \bibfield  {author} {\bibinfo {author} {\bibfnamefont {I.}~\bibnamefont
  {Novikau}}, \bibinfo {author} {\bibfnamefont {E.~A.}\ \bibnamefont
  {Startsev}},\ and\ \bibinfo {author} {\bibfnamefont {I.~Y.}\ \bibnamefont
  {Dodin}},\ }\bibfield  {title} {\bibinfo {title} {Quantum signal processing
  for simulating cold plasma waves},\ }\href
  {https://doi.org/10.1103/PhysRevA.105.062444} {\bibfield  {journal} {\bibinfo
   {journal} {Phys. Rev. A}\ }\textbf {\bibinfo {volume} {105}},\ \bibinfo
  {pages} {062444} (\bibinfo {year} {2022})}\BibitemShut {NoStop}%
\bibitem [{\citenamefont {Vahala}\ \emph {et~al.}(2022)\citenamefont {Vahala},
  \citenamefont {Hawthorne}, \citenamefont {Vahala}, \citenamefont {Ram},\ and\
  \citenamefont {Soe}}]{vahala2022quantum}%
  \BibitemOpen
  \bibfield  {author} {\bibinfo {author} {\bibfnamefont {G.}~\bibnamefont
  {Vahala}}, \bibinfo {author} {\bibfnamefont {J.}~\bibnamefont {Hawthorne}},
  \bibinfo {author} {\bibfnamefont {L.}~\bibnamefont {Vahala}}, \bibinfo
  {author} {\bibfnamefont {A.~K.}\ \bibnamefont {Ram}},\ and\ \bibinfo {author}
  {\bibfnamefont {M.}~\bibnamefont {Soe}},\ }\bibfield  {title} {\bibinfo
  {title} {Quantum lattice representation for the curl equations of maxwell
  equations},\ }\href
  {https://www.tandfonline.com/doi/full/10.1080/10420150.2022.2049784?casa_token=YBLBkPi4__cAAAAA\%3AzJ0f7MHksHlxImG-kfAdKZh7EQlGCNa9Y5kAE_UdbYH0KudzNsJQMvqY-tD2uEPRhjeLCx4YqT4e}
  {\bibfield  {journal} {\bibinfo  {journal} {Radiation Effects and Defects in
  Solids}\ ,\ \bibinfo {pages} {1}} (\bibinfo {year} {2022})}\BibitemShut
  {NoStop}%
\bibitem [{\citenamefont {Engel}\ \emph {et~al.}(2019)\citenamefont {Engel},
  \citenamefont {Smith},\ and\ \citenamefont {Parker}}]{engel2019quantum}%
  \BibitemOpen
  \bibfield  {author} {\bibinfo {author} {\bibfnamefont {A.}~\bibnamefont
  {Engel}}, \bibinfo {author} {\bibfnamefont {G.}~\bibnamefont {Smith}},\ and\
  \bibinfo {author} {\bibfnamefont {S.~E.}\ \bibnamefont {Parker}},\ }\bibfield
   {title} {\bibinfo {title} {Quantum algorithm for the {V}lasov equation},\
  }\href {https://doi.org/10.1103/PhysRevA.100.062315} {\bibfield  {journal}
  {\bibinfo  {journal} {Physical Review A}\ }\textbf {\bibinfo {volume}
  {100}},\ \bibinfo {pages} {062315} (\bibinfo {year} {2019})}\BibitemShut
  {NoStop}%
\bibitem [{\citenamefont {Landau}(1965)}]{landau196561}%
  \BibitemOpen
  \bibfield  {author} {\bibinfo {author} {\bibfnamefont {L.}~\bibnamefont
  {Landau}},\ }\bibfield  {title} {\bibinfo {title} {On the vibrations of the
  electronic plasma},\ }\href
  {https://doi.org/10.1016/B978-0-08-010586-4.50066-3} {\bibfield  {journal}
  {\bibinfo  {journal} {The Collected Papers of LD Landau}\ ,\ \bibinfo {pages}
  {445}} (\bibinfo {year} {1965})}\BibitemShut {NoStop}%
\bibitem [{\citenamefont {Burden}\ \emph {et~al.}(2015)\citenamefont {Burden},
  \citenamefont {Faires},\ and\ \citenamefont {Burden}}]{burden2015numerical}%
  \BibitemOpen
  \bibfield  {author} {\bibinfo {author} {\bibfnamefont {R.~L.}\ \bibnamefont
  {Burden}}, \bibinfo {author} {\bibfnamefont {J.~D.}\ \bibnamefont {Faires}},\
  and\ \bibinfo {author} {\bibfnamefont {A.~M.}\ \bibnamefont {Burden}},\
  }\href@noop {} {\emph {\bibinfo {title} {Numerical analysis}}}\ (\bibinfo
  {publisher} {Cengage learning},\ \bibinfo {year} {2015})\BibitemShut
  {NoStop}%
\bibitem [{\citenamefont {Chen}(2012)}]{chen2012introduction}%
  \BibitemOpen
  \bibfield  {author} {\bibinfo {author} {\bibfnamefont {F.~F.}\ \bibnamefont
  {Chen}},\ }\href@noop {} {\emph {\bibinfo {title} {Introduction to plasma
  physics}}}\ (\bibinfo  {publisher} {Springer Science \& Business Media},\
  \bibinfo {year} {2012})\BibitemShut {NoStop}%
\bibitem [{\citenamefont {Kanekar}\ \emph {et~al.}(2015)\citenamefont
  {Kanekar}, \citenamefont {Schekochihin}, \citenamefont {Dorland},\ and\
  \citenamefont {Loureiro}}]{kanekar2015fluctuation}%
  \BibitemOpen
  \bibfield  {author} {\bibinfo {author} {\bibfnamefont {A.}~\bibnamefont
  {Kanekar}}, \bibinfo {author} {\bibfnamefont {A.}~\bibnamefont
  {Schekochihin}}, \bibinfo {author} {\bibfnamefont {W.}~\bibnamefont
  {Dorland}},\ and\ \bibinfo {author} {\bibfnamefont {N.}~\bibnamefont
  {Loureiro}},\ }\bibfield  {title} {\bibinfo {title} {Fluctuation-dissipation
  relations for a plasma-kinetic {L}angevin equation},\ }\href
  {https://doi.org/10.1017/S0022377814000622} {\bibfield  {journal} {\bibinfo
  {journal} {Journal of Plasma Physics}\ }\textbf {\bibinfo {volume} {81}}
  (\bibinfo {year} {2015})}\BibitemShut {NoStop}%
\bibitem [{\citenamefont {Adkins}\ and\ \citenamefont
  {Schekochihin}(2018)}]{adkins2018solvable}%
  \BibitemOpen
  \bibfield  {author} {\bibinfo {author} {\bibfnamefont {T.}~\bibnamefont
  {Adkins}}\ and\ \bibinfo {author} {\bibfnamefont {A.}~\bibnamefont
  {Schekochihin}},\ }\bibfield  {title} {\bibinfo {title} {A solvable model of
  {V}lasov-kinetic plasma turbulence in {F}ourier--{H}ermite phase space},\
  }\href
  {https://www.cambridge.org/core/journals/journal-of-plasma-physics/article/abs/solvable-model-of-vlasovkinetic-plasma-turbulence-in-fourierhermite-phase-space/000E7BCAA305329FD131F8A9245ABB6C}
  {\bibfield  {journal} {\bibinfo  {journal} {Journal of Plasma Physics}\
  }\textbf {\bibinfo {volume} {84}} (\bibinfo {year} {2018})}\BibitemShut
  {NoStop}%
\bibitem [{\citenamefont {Armstrong}(1967)}]{armstrong1967numerical}%
  \BibitemOpen
  \bibfield  {author} {\bibinfo {author} {\bibfnamefont {T.~P.}\ \bibnamefont
  {Armstrong}},\ }\bibfield  {title} {\bibinfo {title} {Numerical studies of
  the nonlinear {V}lasov equation},\ }\href {https://doi.org/10.1063/1.1762272}
  {\bibfield  {journal} {\bibinfo  {journal} {The Physics of Fluids}\ }\textbf
  {\bibinfo {volume} {10}},\ \bibinfo {pages} {1269} (\bibinfo {year}
  {1967})}\BibitemShut {NoStop}%
\bibitem [{\citenamefont {Grant}\ and\ \citenamefont
  {Feix}(1967)}]{grant1967fourier}%
  \BibitemOpen
  \bibfield  {author} {\bibinfo {author} {\bibfnamefont {F.~C.}\ \bibnamefont
  {Grant}}\ and\ \bibinfo {author} {\bibfnamefont {M.~R.}\ \bibnamefont
  {Feix}},\ }\bibfield  {title} {\bibinfo {title} {Fourier-{H}ermite solutions
  of the {V}lasov equations in the linearized limit},\ }\href
  {https://doi.org/10.1063/1.1762177} {\bibfield  {journal} {\bibinfo
  {journal} {The Physics of Fluids}\ }\textbf {\bibinfo {volume} {10}},\
  \bibinfo {pages} {696} (\bibinfo {year} {1967})}\BibitemShut {NoStop}%
\bibitem [{\citenamefont {Armstrong}\ \emph {et~al.}(1970)\citenamefont
  {Armstrong}, \citenamefont {Harding}, \citenamefont {Knorr},\ and\
  \citenamefont {Montgomery}}]{armstrong1970solution}%
  \BibitemOpen
  \bibfield  {author} {\bibinfo {author} {\bibfnamefont {T.~P.}\ \bibnamefont
  {Armstrong}}, \bibinfo {author} {\bibfnamefont {R.~C.}\ \bibnamefont
  {Harding}}, \bibinfo {author} {\bibfnamefont {G.}~\bibnamefont {Knorr}},\
  and\ \bibinfo {author} {\bibfnamefont {D.}~\bibnamefont {Montgomery}},\
  }\bibfield  {title} {\bibinfo {title} {Solution of {V}lasov equation by
  transform methods},\ }\href@noop {} {\bibfield  {journal} {\bibinfo
  {journal} {Methods of Computational Physics}\ }\textbf {\bibinfo {volume}
  {9}},\ \bibinfo {pages} {29} (\bibinfo {year} {1970})}\BibitemShut {NoStop}%
\bibitem [{\citenamefont {Hammett}\ \emph {et~al.}(1993)\citenamefont
  {Hammett}, \citenamefont {Beer}, \citenamefont {Dorland}, \citenamefont
  {Cowley},\ and\ \citenamefont {Smith}}]{hammett1993developments}%
  \BibitemOpen
  \bibfield  {author} {\bibinfo {author} {\bibfnamefont {G.}~\bibnamefont
  {Hammett}}, \bibinfo {author} {\bibfnamefont {M.}~\bibnamefont {Beer}},
  \bibinfo {author} {\bibfnamefont {W.}~\bibnamefont {Dorland}}, \bibinfo
  {author} {\bibfnamefont {S.}~\bibnamefont {Cowley}},\ and\ \bibinfo {author}
  {\bibfnamefont {S.}~\bibnamefont {Smith}},\ }\bibfield  {title} {\bibinfo
  {title} {Developments in the gyrofluid approach to tokamak turbulence
  simulations},\ }\href {https://doi.org/10.1088/0741-3335/35/8/006} {\bibfield
   {journal} {\bibinfo  {journal} {Plasma physics and controlled fusion}\
  }\textbf {\bibinfo {volume} {35}},\ \bibinfo {pages} {973} (\bibinfo {year}
  {1993})}\BibitemShut {NoStop}%
\bibitem [{\citenamefont {Parker}\ and\ \citenamefont
  {Carati}(1995)}]{parker1995renormalized}%
  \BibitemOpen
  \bibfield  {author} {\bibinfo {author} {\bibfnamefont {S.~E.}\ \bibnamefont
  {Parker}}\ and\ \bibinfo {author} {\bibfnamefont {D.}~\bibnamefont
  {Carati}},\ }\bibfield  {title} {\bibinfo {title} {Renormalized dissipation
  in plasmas with finite collisionality},\ }\href
  {https://doi.org/10.1103/PhysRevLett.75.441} {\bibfield  {journal} {\bibinfo
  {journal} {Physical Review Letters}\ }\textbf {\bibinfo {volume} {75}},\
  \bibinfo {pages} {441} (\bibinfo {year} {1995})}\BibitemShut {NoStop}%
\bibitem [{\citenamefont {Sugama}\ \emph {et~al.}(2001)\citenamefont {Sugama},
  \citenamefont {Watanabe},\ and\ \citenamefont
  {Horton}}]{sugama2001collisionless}%
  \BibitemOpen
  \bibfield  {author} {\bibinfo {author} {\bibfnamefont {H.}~\bibnamefont
  {Sugama}}, \bibinfo {author} {\bibfnamefont {T.-H.}\ \bibnamefont
  {Watanabe}},\ and\ \bibinfo {author} {\bibfnamefont {W.}~\bibnamefont
  {Horton}},\ }\bibfield  {title} {\bibinfo {title} {Collisionless
  kinetic-fluid closure and its application to the three-mode ion temperature
  gradient driven system},\ }\href {https://doi.org/10.1063/1.1367319}
  {\bibfield  {journal} {\bibinfo  {journal} {Physics of Plasmas}\ }\textbf
  {\bibinfo {volume} {8}},\ \bibinfo {pages} {2617} (\bibinfo {year}
  {2001})}\BibitemShut {NoStop}%
\bibitem [{\citenamefont {Zocco}\ and\ \citenamefont
  {Schekochihin}(2011)}]{zocco2011reduced}%
  \BibitemOpen
  \bibfield  {author} {\bibinfo {author} {\bibfnamefont {A.}~\bibnamefont
  {Zocco}}\ and\ \bibinfo {author} {\bibfnamefont {A.~A.}\ \bibnamefont
  {Schekochihin}},\ }\bibfield  {title} {\bibinfo {title} {Reduced
  fluid-kinetic equations for low-frequency dynamics, magnetic reconnection,
  and electron heating in low-beta plasmas},\ }\href
  {https://doi.org/10.1063/1.3628639} {\bibfield  {journal} {\bibinfo
  {journal} {Physics of Plasmas}\ }\textbf {\bibinfo {volume} {18}},\ \bibinfo
  {pages} {102309} (\bibinfo {year} {2011})}\BibitemShut {NoStop}%
\bibitem [{\citenamefont {Lenard}\ and\ \citenamefont
  {Bernstein}(1958)}]{lenard1958plasma}%
  \BibitemOpen
  \bibfield  {author} {\bibinfo {author} {\bibfnamefont {A.}~\bibnamefont
  {Lenard}}\ and\ \bibinfo {author} {\bibfnamefont {I.~B.}\ \bibnamefont
  {Bernstein}},\ }\bibfield  {title} {\bibinfo {title} {Plasma oscillations
  with diffusion in velocity space},\ }\href
  {https://doi.org/10.1103/PhysRev.112.1456} {\bibfield  {journal} {\bibinfo
  {journal} {Physical Review}\ }\textbf {\bibinfo {volume} {112}},\ \bibinfo
  {pages} {1456} (\bibinfo {year} {1958})}\BibitemShut {NoStop}%
\bibitem [{\citenamefont {S{\"u}nderhauf}\ \emph {et~al.}(2023)\citenamefont
  {S{\"u}nderhauf}, \citenamefont {Campbell},\ and\ \citenamefont
  {Camps}}]{sunderhauf2023block}%
  \BibitemOpen
  \bibfield  {author} {\bibinfo {author} {\bibfnamefont {C.}~\bibnamefont
  {S{\"u}nderhauf}}, \bibinfo {author} {\bibfnamefont {E.}~\bibnamefont
  {Campbell}},\ and\ \bibinfo {author} {\bibfnamefont {J.}~\bibnamefont
  {Camps}},\ }\bibfield  {title} {\bibinfo {title} {Block-encoding structured
  matrices for data input in quantum computing},\ }\href
  {https://doi.org/10.48550/arXiv.2302.10949} {\bibfield  {journal} {\bibinfo
  {journal} {arXiv preprint arXiv:2302.10949}\ } (\bibinfo {year}
  {2023})}\BibitemShut {NoStop}%
\bibitem [{\citenamefont {Brassard}\ \emph {et~al.}(2002)\citenamefont
  {Brassard}, \citenamefont {Hoyer}, \citenamefont {Mosca},\ and\ \citenamefont
  {Tapp}}]{brassard2002quantum}%
  \BibitemOpen
  \bibfield  {author} {\bibinfo {author} {\bibfnamefont {G.}~\bibnamefont
  {Brassard}}, \bibinfo {author} {\bibfnamefont {P.}~\bibnamefont {Hoyer}},
  \bibinfo {author} {\bibfnamefont {M.}~\bibnamefont {Mosca}},\ and\ \bibinfo
  {author} {\bibfnamefont {A.}~\bibnamefont {Tapp}},\ }\bibfield  {title}
  {\bibinfo {title} {Quantum amplitude amplification and estimation},\ }\href
  {https://doi.org/10.1090/conm/305/05215} {\bibfield  {journal} {\bibinfo
  {journal} {Contemporary Mathematics}\ }\textbf {\bibinfo {volume} {305}},\
  \bibinfo {pages} {53} (\bibinfo {year} {2002})}\BibitemShut {NoStop}%
\bibitem [{Note1()}]{Note1}%
  \BibitemOpen
  \bibinfo {note} {The linear Vlasov equation exhibits a phenomenon known as
  recursion, where, for a long simulation time, the velocity space cascade
  towards finer structures reverses. This phenomenon is purely numerical and
  requires one to choose the velocity space resolution to be high enough (or
  the simulation time to be small enough) to prevent it. For the error
  analysis, we will assume that this is indeed the case, so as to avoid the
  problem of recursion.}\BibitemShut {Stop}%
\bibitem [{\citenamefont {Boyd}(1984)}]{boyd1984asymptotic}%
  \BibitemOpen
  \bibfield  {author} {\bibinfo {author} {\bibfnamefont {J.~P.}\ \bibnamefont
  {Boyd}},\ }\bibfield  {title} {\bibinfo {title} {Asymptotic coefficients of
  {H}ermite function series},\ }\href
  {https://doi.org/10.1016/0021-9991(84)90124-4} {\bibfield  {journal}
  {\bibinfo  {journal} {Journal of Computational Physics}\ }\textbf {\bibinfo
  {volume} {54}},\ \bibinfo {pages} {382} (\bibinfo {year} {1984})}\BibitemShut
  {NoStop}%
\bibitem [{\citenamefont {Boyd}(2001)}]{boyd2001chebyshev}%
  \BibitemOpen
  \bibfield  {author} {\bibinfo {author} {\bibfnamefont {J.~P.}\ \bibnamefont
  {Boyd}},\ }\href@noop {} {\emph {\bibinfo {title} {Chebyshev and {F}ourier
  spectral methods}}}\ (\bibinfo  {publisher} {Courier Corporation},\ \bibinfo
  {year} {2001})\BibitemShut {NoStop}%
\bibitem [{\citenamefont {Rall}(1965)}]{rall1965numerical}%
  \BibitemOpen
  \bibfield  {author} {\bibinfo {author} {\bibfnamefont {L.}~\bibnamefont
  {Rall}},\ }\bibfield  {title} {\bibinfo {title} {Numerical integration and
  the solution of integral equations by the use of {R}iemann sums},\ }\href
  {https://doi.org/10.1137/1007005} {\bibfield  {journal} {\bibinfo  {journal}
  {SIAM Review}\ }\textbf {\bibinfo {volume} {7}},\ \bibinfo {pages} {55}
  (\bibinfo {year} {1965})}\BibitemShut {NoStop}%
\bibitem [{\citenamefont {Kitaev}(1995)}]{kitaev1995quantum}%
  \BibitemOpen
  \bibfield  {author} {\bibinfo {author} {\bibfnamefont {A.~Y.}\ \bibnamefont
  {Kitaev}},\ }\bibfield  {title} {\bibinfo {title} {Quantum measurements and
  the {A}belian stabilizer problem},\ }\href
  {https://doi.org/10.48550/arXiv.quant-ph/9511026} {\bibfield  {journal}
  {\bibinfo  {journal} {arXiv preprint quant-ph/9511026}\ } (\bibinfo {year}
  {1995})}\BibitemShut {NoStop}%
\bibitem [{\citenamefont {Nielsen}\ and\ \citenamefont
  {Chuang}(2010)}]{nielsen2002quantum}%
  \BibitemOpen
  \bibfield  {author} {\bibinfo {author} {\bibfnamefont {M.}~\bibnamefont
  {Nielsen}}\ and\ \bibinfo {author} {\bibfnamefont {I.}~\bibnamefont
  {Chuang}},\ }\href@noop {} {\emph {\bibinfo {title} {Quantum Computation and
  Quantum Information: 10th Anniversary Edition}}}\ (\bibinfo  {publisher}
  {Cambridge University Press},\ \bibinfo {year} {2010})\BibitemShut {NoStop}%
\bibitem [{\citenamefont {Abrams}\ and\ \citenamefont
  {Lloyd}(1999)}]{abrams1999quantum}%
  \BibitemOpen
  \bibfield  {author} {\bibinfo {author} {\bibfnamefont {D.~S.}\ \bibnamefont
  {Abrams}}\ and\ \bibinfo {author} {\bibfnamefont {S.}~\bibnamefont {Lloyd}},\
  }\bibfield  {title} {\bibinfo {title} {Quantum algorithm providing
  exponential speed increase for finding eigenvalues and eigenvectors},\ }\href
  {https://doi.org/10.1103/PhysRevLett.83.5162} {\bibfield  {journal} {\bibinfo
   {journal} {Physical Review Letters}\ }\textbf {\bibinfo {volume} {83}},\
  \bibinfo {pages} {5162} (\bibinfo {year} {1999})}\BibitemShut {NoStop}%
\bibitem [{\citenamefont {Jaksch}\ and\ \citenamefont
  {Papageorgiou}(2003)}]{jaksch2003eigenvector}%
  \BibitemOpen
  \bibfield  {author} {\bibinfo {author} {\bibfnamefont {P.}~\bibnamefont
  {Jaksch}}\ and\ \bibinfo {author} {\bibfnamefont {A.}~\bibnamefont
  {Papageorgiou}},\ }\bibfield  {title} {\bibinfo {title} {Eigenvector
  approximation leading to exponential speedup of quantum eigenvalue
  calculation},\ }\href {https://doi.org/10.1103/PhysRevLett.91.257902}
  {\bibfield  {journal} {\bibinfo  {journal} {Physical review letters}\
  }\textbf {\bibinfo {volume} {91}},\ \bibinfo {pages} {257902} (\bibinfo
  {year} {2003})}\BibitemShut {NoStop}%
\bibitem [{\citenamefont {Wang}\ \emph {et~al.}(2010)\citenamefont {Wang},
  \citenamefont {Wu}, \citenamefont {Liu},\ and\ \citenamefont
  {Nori}}]{wang2010measurement}%
  \BibitemOpen
  \bibfield  {author} {\bibinfo {author} {\bibfnamefont {H.}~\bibnamefont
  {Wang}}, \bibinfo {author} {\bibfnamefont {L.-A.}\ \bibnamefont {Wu}},
  \bibinfo {author} {\bibfnamefont {Y.-x.}\ \bibnamefont {Liu}},\ and\ \bibinfo
  {author} {\bibfnamefont {F.}~\bibnamefont {Nori}},\ }\bibfield  {title}
  {\bibinfo {title} {Measurement-based quantum phase estimation algorithm for
  finding eigenvalues of non-unitary matrices},\ }\href
  {https://doi.org/10.1103/PhysRevA.82.062303} {\bibfield  {journal} {\bibinfo
  {journal} {Physical Review A}\ }\textbf {\bibinfo {volume} {82}},\ \bibinfo
  {pages} {062303} (\bibinfo {year} {2010})}\BibitemShut {NoStop}%
\bibitem [{\citenamefont {Daskin}\ \emph {et~al.}(2014)\citenamefont {Daskin},
  \citenamefont {Grama},\ and\ \citenamefont {Kais}}]{daskin2014universal}%
  \BibitemOpen
  \bibfield  {author} {\bibinfo {author} {\bibfnamefont {A.}~\bibnamefont
  {Daskin}}, \bibinfo {author} {\bibfnamefont {A.}~\bibnamefont {Grama}},\ and\
  \bibinfo {author} {\bibfnamefont {S.}~\bibnamefont {Kais}},\ }\bibfield
  {title} {\bibinfo {title} {A universal quantum circuit scheme for finding
  complex eigenvalues},\ }\href {https://doi.org/10.1007/s11128-013-0654-1}
  {\bibfield  {journal} {\bibinfo  {journal} {Quantum information processing}\
  }\textbf {\bibinfo {volume} {13}},\ \bibinfo {pages} {333} (\bibinfo {year}
  {2014})}\BibitemShut {NoStop}%
\bibitem [{\citenamefont {Shao}(2022)}]{shao2022computing}%
  \BibitemOpen
  \bibfield  {author} {\bibinfo {author} {\bibfnamefont {C.}~\bibnamefont
  {Shao}},\ }\bibfield  {title} {\bibinfo {title} {Computing eigenvalues of
  diagonalizable matrices on a quantum computer},\ }\href
  {https://doi.org/10.1145/3527845} {\bibfield  {journal} {\bibinfo  {journal}
  {ACM Transactions on Quantum Computing}\ }\textbf {\bibinfo {volume} {3}},\
  \bibinfo {pages} {1} (\bibinfo {year} {2022})}\BibitemShut {NoStop}%
\end{thebibliography}%

\end{document}